\documentclass{article}

\usepackage[round]{natbib}
\usepackage{amsmath,amssymb,amsfonts}
\usepackage{algorithmic}
\usepackage[labelformat=simple]{subfig}

\usepackage{graphicx}
\usepackage{textcomp}
\usepackage{xcolor}
\usepackage{bm}
\usepackage{pdflscape}
\usepackage{rotating}
\usepackage{array}
\usepackage{tabularx,booktabs}
\usepackage{amsmath}
\usepackage{url}
\usepackage[flushleft]{threeparttable}

\usepackage{geometry}

\usepackage{makecell}

\newcommand{\Rc}{\mathcal{R}}
\newcommand{\R}{\mathbb{R}}

\renewcommand{\P}{\mathbb{P}}

\newcommand{\Tau}{{\mathcal{T}}}

\newcommand{\diag}{\textrm{\rm diag}}

\newcommand{\VaR}{\textrm{\rm VaR}}
\newcommand{\CVaR}{\textrm{\rm CVaR}}

\newcommand{\LR}[1]{\left({#1}\right)}

\newcolumntype{C}{>{\centering\arraybackslash}X} %

\title{Portfolio Optimization on Multivariate Regime-Switching
GARCH Model with Normal Tempered Stable Innovation}

\date{}

\author{Cheng Peng\thanks{Stony Brook University, cheng.peng.1@stonybrook.edu}
\,
 Young Shin Kim\thanks{Stony Brook University, aaron.Kim@stonybrook.edu}
\,
and Stefan Mittnik\thanks{
Ludwig Maximilians University in Munich, finmetrics@stat.uni-muenchen.de}
}

\begin{document}

\maketitle

\begin{abstract}

This paper uses simulation-based portfolio optimization to mitigate the left tail risk of the portfolio. The contribution is twofold. (i) We propose the Markov regime-switching GARCH model with multivariate normal tempered stable innovation (MRS-MNTS-GARCH) to accommodate fat tails, volatility clustering and regime switch. The volatility of each asset independently follows the regime-switch  GARCH model, while the correlation of joint innovation of the GARCH models follows the Hidden Markov Model.  (ii) We  use tail risk measures, namely conditional value-at-risk (CVaR) and conditional drawdown-at-risk (CDaR), in the portfolio optimization. The optimization is performed  with the sample paths simulated by the MRS-MNTS-GARCH model. We conduct an empirical study on the performance of optimal portfolios.
Out-of-sample tests show that the optimal portfolios with tail measures outperform the optimal portfolio with standard deviation measure and the equally weighted portfolio in various performance measures. The 
out-of-sample performance of the optimal portfolios is also more robust to suboptimality 
on the efficient frontier.
\end{abstract}

\section{Introduction}

Empirical studies have found in the return of various financial instruments skewness and leptokurtotic—asymmetry, and higher peak around the mean with fat tails. Normal distribution has long been recognized as insufficient to accommodate these stylized facts, relying on which could drastically underestimate the tail risk of a portfolio. The $\alpha$-stable distribution is a candidate to incorporate this fact into decision-making, but its lack of finite moments could cause difficulties in modeling. The class of tempered stable distributions serves as a natural substitution, since it has finite moments while retaining many attractive properties of the  $\alpha$-stable distribution. The class of tempered stable distributions is derived by tempering the tails of the $\alpha$-stable distribution  (See \citeauthor{Koponen:1995} \citeyear{Koponen:1995}; \citeauthor{Boyarchenko_Levendorskii:2000} \citeyear{Boyarchenko_Levendorskii:2000}; and \citeauthor{CGMY:2002} \citeyear{CGMY:2002}), or from the time changed Brownian motion. Since \cite{Barndorff-Nielsen1} and \cite{Barndorff-Nielsen2} presented the normal tempered stable (NTS) distribution in finance using the time changed Brownian motion, it has been successfully applied for modeling the stock returns with high accuracy in literature including \cite{Kim2}, \cite{Kim4}, \cite{Kim5} and \cite{Kim6}.

Deviation from normality exists not only in raw returns, but also in  filtered innovations of time series models. Since the foundational work in \cite{Engle1} and \cite{Bollerslev}, GARCH model has been widely applied to model volatility. This motivates us to use  NTS distribution to accommodate the asymmetry, interdependence, and fat tails of the innovations of GARCH models. Several studies have found NTS distribution to be more favorable than other candidates. \cite{Kim1} found that normal and $\textit{t}$-distribution are rejected as innovation of GARCH model in hypothesis testing. \cite{Shi} found that tempered stable distribution with exponential tempering yields better calibration of Markov Regime-Switching (MRS) GARCH model than $\textit{t}$ or generalized error distribution.  Generalized hyperbolic distribution, while very flexible, has too many parameters for estimation. NTS distribution has only three standardized parameters and is flexible enough to serve the purpose.

A drawback of GARCH model is the inadequate ability in modeling volatility spikes, which could indicate that the market is switching within regimes. Various types of  MRS-GARCH models have been proposed, many of which suggest that the regime-switching GARCH model achieves a better fit for empirical data (for example, in \citeauthor{Hamilton} \citeyear{Hamilton}). 
\cite{Marcucci} finds that MRS-GARCH models outperform standard GARCH models in volatility forecasting in a time horizon shorter than one week. Naturally, the most flexible model allows all parameters to switch among regimes. However, as is shown in \mbox{\cite{Henneke}}, the sampling procedure in the MCMC method for estimating such a model is time-consuming and renders it improper for practical use. We construct a new model based on the regime-switching GARCH model specified in \cite{Haas1}, which circumvents the path dependence problem in the Markov Chain model by specifying parallel GARCH models.

\cite{Bollerslev2} proposes the GARCH model with constant conditional correlation. The parsimonious DCC-GARCH model in \cite{Engle2} opens the door to modeling multivariate GARCH process with different persistence between variance and correlation. While flexible models allowing time-varying correlation are preferred in many cases, specification of multivariate GARCH models with both regime-specific correlations and variance dynamics involves a balance between flexibility and tractability. The univariate MRS-GARCH model in \cite{Haas1} is generalized in \cite{Haas2} to a multivariate case. Unfortunately, it suffers from the curse of dimensionality, and thus is unsuitable for a high dimensional empirical study. In our model, we decompose variance and correlation so that the variance of each asset evolves independently according to a univariate MRS-GARCH model. The correlation is incorporated in the innovations modeled by a flexible Hidden Markov Model (HMM) that has multivariate NTS distribution as the conditional distribution in each regime.  We will use the calibrated model to simulate returns for portfolio~optimizations. 

Modern portfolio theory is formulated as a trade-off between return and risk. The classical Markowitz Model finds the portfolio with the highest Sharpe ratio. However, variance has been criticized for not containing enough information on the tail of the distribution. Moreover, correlation is not sufficient to describe the interdependence of asset returns with non-elliptical distribution. 
Current regulations for financial business utilize the concept of Value at Risk (VaR), which is the percentile of the loss distribution,  to model the risk of left tail events. There are several undesired properties that rendered it an insufficient criterion. First, it is not a coherent risk measure due to a lack of sub-additivity. Second, VaR is a non-convex and non-smooth  function of positions with multiple local extrema for non-normal distributions, which causes difficulty in developing efficient optimization techniques. Third, a single percentile is insufficient to describe the tail behavior of a distribution, and thus might lead to an underestimation of risk. 

 Theory and algorithm for portfolio optimization with a conditional value-at-risk (CVaR) measure proposed in \cite{Uryasev5} and \cite{Uryasev6} effectively addresses these problems. For continuous distributions, CVaR is defined by a conditional expectation of losses exceeding a certain VaR level. For discrete distributions, CVaR is defined by a weighted average of some VaR levels that exceed a specified VaR level. In this way, CVaR concerns both VaR and the losses exceeding VaR.  As a convex function of asset allocation, a coherent risk measure and a more informative statistic, CVaR serves as an ideal alternative to VaR. A study on the comparison of the two measures can be found in \cite{uryasev7}.

Another closely related tail risk measure is conditional drawdown-at-risk (CDaR). Drawdown has been a constant concern for investors and is often used in the performance  evaluation of a portfolio.   It is much more difficult to climb out of a drawdown than drop into one, considering that it takes 100\% return to recover from 50\% relative drawdown.   In behavioral finance, it is well documented that investors fear losses more than they value gains. 
However, the commonly used maximum drawdown only considers the worst scenario that only occur under some very special circumstances, and thus is very sensitive to the testing period and asset allocation. On the other hand, small drawdowns included in the calculation of average drawdown are acceptable and might be caused by pure noise, of which the minimization might result in overfitting. For instance, a Brownian motion would have drawdowns in a simulated sample path. 

CDaR is proposed in \cite{Uryasev2} to address these concerns. While CVaR is determined by the distribution of return, CDaR is path-dependent. CDaR is essentially CVaR of the distribution of drawdowns. By this, we overcome the drawbacks of average drawdown and maximum drawdown. CDaR takes not only the depth of drawdowns into consideration, but also the length of them.  Since the CDaR risk measure is the application of CVaR in a dynamic case, it naturally holds nice properties of CVaR, such as convexity with respect to asset allocation. The optimization method with constraints on CDaR has also been developed and studied in \citeauthor{Uryasev2} (\citeyear{Uryasev2}, \citeyear{Uryasev1}). 

An optimization procedure could lead to an underperformed portfolio by magnifying the estimation error. Using historical sample paths as input 
implies an assumption that what happened in the past will happen in the future, which requires careful examination. It is found in \cite{Lim} that estimation errors of CVaR and the mean is large, resulting in unreliable allocation. An alternative way is to use multiple simulated returns. It is reasonable to expect outperformance during a crisis if we use the simulation of a model that captures extreme tail events as input for optimization with tail risk measures. 
Relevant research that resorts to copula does not address a regime switch (See
\mbox{\citeauthor{Sahamkhadam} \citeyear{Sahamkhadam}}).

We have identified several issues in GARCH-simulation-based portfolio optimization: deviation from normality, regime switch, high dimensional calibration and tail risk measures. This  paper intends to address these issues in one shot. First, we propose the MRS-MNTS-GARCH model to accommodate fat tails, volatility clustering and regime switch. In the model, the volatility of each asset independently follows a regime-switch  GARCH model, while the correlation of joint innovation of GARCH models follows a Hidden Markov Model. We specify the method for efficient model calibration. The MRS-MNTS-GARCH model is used to simulate sample paths. Then, the sample paths are used as input to portfolio optimization with tail risk measures CVaR and CDaR. We conduct in-sample tests sow show the goodness-of-fit. We conduct out-of-sample tests to show the outperformance of our approach in various measures. The optimal portfolios are also more robust to suboptimality of the optimization. The results suggest that, in practice, performance ratios can be improved compared to equally weighted portfolio by adopting simulation-based portfolio optimization with the MRS-MNTS-GARCH model. The empirical study suggests that using tail risk measures outperforms using standard deviation in terms of various performance measures and robustness.

The remainder of the paper is organized as follows. Section \ref{sec_Preliminaries} introduces the preliminaries on NTS distribution and GARCH model. Section \ref{sec_Model} specifies our model and methods for estimation and simulation. Section \ref{sec_Portfolio_Optimization} introduces the portfolio optimization with tail risk measures.  Section \ref{sec_Empirical_Study} is an empirical study on in-sample goodness-of-fit and out-of-sample performance in recent years.

\section{Preliminaries} \label{sec_Preliminaries}

This section reviews the background knowledge on different risk measures, scenario-based estimation, normal tempered stable (NTS) distribution, and GARCH and regime-switching GARCH model.

\subsection{Tail Risk Measures}
In this subsection, we discuss and summarize popular tail risk measures in finance such as Value at Risk (VaR), Conditional VaR (CVaR), and conditional drawdown-at-risk (CDaR). Definitions of CDaR, and related properties of drawdown are following \mbox{\citeauthor{Uryasev2} (\citeyear{Uryasev2}, \citeyear{Uryasev1})}.

Consider a probability space $(\Omega, \mathcal F_\infty, \P)$ and a portfolio with $N$ assets.
Suppose that $P_n$ is a stochastic process of a price for the $n$-th asset in the portfolio on the space 
\[
P_n:[0,\infty)\times \Omega \longrightarrow \R, ~~~ n = 1,2,\cdots, N
\]
with the real initial value $P_n(0,\cdot) = P_{n,0}>0$.

The stochastic process of the compounded return of the $n$-th asset between time 0 to time $t>0$ is defined by
\[
\Rc_n(t, \omega) = \frac{P_n(t,\omega)-P_{n,0}}{P_{n,0}}, ~~~ t>0,
\]
with $\Rc_n(0, \cdot)=0$. 
Then, the compounded portfolio return $R(x, t,\omega)$ at time $t>0$ is equal~to
\[
\Rc(x, t, \omega) = \sum_{n=1}^N x_n \Rc_n(t,\omega).
\]

The definition of VaR for $\Rc(x, t, \cdot)$ with the confidence level $\eta$ is
\[
\VaR_\eta( \Rc(x, t, \cdot))=\min \{ z| \P( \Rc(x, t, \cdot)\le -z) \geq \eta\}.
\]

If $\Rc(x, t, \cdot)$ has a continuous cumulative distribution, then we have
\[
\VaR_\eta(\Rc(x, t, \cdot))=F_{-\Rc(x, t, \cdot)}^{-1}(1-\eta),
\]
where $F_{-\Rc(x, t, \cdot)}^{-1}$ is the inverse cumulative distribution function of $-\Rc(x, t, \cdot)$. The definition of CVaR for $\Rc(x, t, \cdot)$ with the confidence level $\eta$ is
\begin{align*}
\CVaR_\eta(\Rc(x, t, \cdot)) = \frac{1}{1-\eta}\int_{\eta}^1
\VaR_{z}(\Rc(x, t, \cdot))dz.
\end{align*}

Equivalently, CVaR can be obtained by a minimization formula
\begin{align}\label{cvar_formula}
\CVaR_\eta(\Rc(x, t, \cdot)) = \min_\theta \left\{ \theta + \frac{1}{\eta}E\left[\max\{0,-\Rc(x, t, \cdot)-\theta\}\right] \right\}.
\end{align}

Following \cite{Uryasev2}, denote the uncompounded return of the $n$-th asset by $\mathcal{Q}_n(t,w)$. The uncompounded portfolio return is defined by
\[
\mathcal{Q}(x, t, \omega) = \sum_{n=1}^N x_n \mathcal{Q}_n(t,\omega).
\]

Let $T>0$, and $\omega\in\Omega$, then $(R(x, t,\omega))_{t\in[0,T]}=\{R(x, t,\omega)|t\in[0,T]\}$ is one sample path of portfolio return with the capital allocation vector $x$ from time 0 to $T$.
The drawdown $(\textup{DD}(\cdot,\omega))_{t\in[0,T]}$ of the portfolio return is defined by
\begin{equation}\label{eq:DD}
\textup{DD}(t, \omega) = \max_{s\in[0,t]} \mathcal{Q}(x, s,\omega) - \mathcal{Q}(x, t,\omega), \text{ for } t\in[0,T]
\end{equation}
is a risk measure assessing the decline from a historical peak in some variable, and $\{\textup{DD}(t,\omega)|t\in[0,T], \omega\in\Omega\}$ is a stochastic process of the drawdown.
The average drawdown (ADD) up to time $T$ for $\omega\in \Omega$ is defined by
\begin{equation}\label{eq:ADD}
\textup{ADD}(T,\omega) = \frac{1}{T}\int_0^T \textup{DD}(t, \omega)dt .
\end{equation}

It is the time average of drawdowns observed between time 0 to $T$.
The maximum drawdown (ADD) up to time $T$ for $\omega\in \Omega$ is defined by
\begin{equation}\label{eq:MDD}
\textup{MDD}(T,\omega) = \max_{t\in(0,T)} \textup{DD}(t, \omega).
\end{equation}

It is the maximum of drawdowns that have occurred up to time $T$ for $\omega\in \Omega$.

Let 
\begin{equation}\label{eq:FDD}
F_{\textup{DD}}(z,\omega) = \frac{1}{T}\int_0^T 1_{\textup{DD}(t,\omega)\le z}dt
\end{equation}
and
\begin{equation}\label{eq:zeta}
\zeta_\eta(\omega) = \begin{cases}
\inf\{z | F_{\textup{DD}}(z,\omega)\ge \eta\} &\text{ if } \eta\in(0,1]\\
0 &\text{ if } \eta=0
\end{cases}.
\end{equation}

According to \cite{Uryasev1}, the CDaR is defined as the CVaR of drawdowns. Applying \eqref{cvar_formula} to drawdowns, we have
\[
\textup{CDaR}_\eta(T, \omega) = \min_{\zeta}\left\{\zeta+\frac{1}{\eta T}\int_0^T \max\left\{\textup{DD}(t,\omega)-\zeta, 0\right\}dt\right\} .
\]

It can also be represented as
\begin{align}
\nonumber
&\textup{CDaR}_\eta(T, \omega) 
\\
&
\label{eq:CDaR}
= \left(\frac{F_{\textup{DD}}(\zeta_\eta(\omega), \omega)-\eta}{1-\eta}\right)\zeta_\eta(\omega)+\frac{1}{(1-\eta) T}\int_0^T \textup{DD}(t, \omega) 1_{\textup{DD}(t,\omega)>\zeta_\eta(\omega)}dt.
\end{align}

Note that $\textup{CDaR}_0(T,\omega)=\textup{ADD}_\omega(T)$ and  $\textup{CDaR}_1(T,\omega)=\textup{MDD}_\omega(T)$.

Let 
\begin{equation}\label{eq:FDD_multi}
F_{\textup{DD}}(z) = \frac{1}{T}E\left[\int_0^T 1_{DD(t,\cdot)\le z}dt\right]
\end{equation}
and 
\begin{equation}\label{eq:zeta_multi}
\zeta_\eta = F_{\textup{DD}}^{-1}(\eta) = \begin{cases}
\inf\{ z|F_{\textup{DD}}(z)\ge\eta\} &\text{ if } \eta\in(0,1]\\
0 &\text{ if } \eta=0
\end{cases}.
\end{equation}

Considering all the sample path $\{\Rc(x, t,\omega) | t\in[0,T], \omega\in\Omega\}$, we define CDaR at $\eta$~by
\begin{align}
\nonumber &\textup{CDaR}_\eta(T) \\
&= \left(\frac{F_{\textup{DD}}(\zeta(\eta))-\eta}{1-\eta}\right)\zeta(\eta)+\frac{1}{(1-\eta) T}E\left[\int_0^T \textup{DD}(t,\cdot)1_{\textup{DD}(t,\cdot)>\zeta(\eta)}dt\right].
\label{eq:CDaR_multi}
\end{align}
\subsection{Scenario-Based Estimation}

In this subsection, we discuss the risk estimation using given scenarios (historical scenario or simulation). Suppose that the time interval $[0,T]$ is divided by a partition $\{ 0=t_0 < t_1 < \cdots < t_M=T\}$, and denote 
\[
R_m(x) = \Rc(x,t_m,\cdot).
\]

We select $\omega_s\in\Omega$ where $s\in\{1,2,\cdots, S\}$ and $S$ is the number of scenarios. Then, we obtain scenarios of the portfolio return process $\Rc$ at time $t_m$: 
\[
R_m^s(x)=\Rc(x,t_m,\omega_s)
\]
where $m\in\{0, 1,2,\cdots, M\}$, and $s\in\{1,2,\cdots, S\}$.
We calculate VaR, CVaR, DD, and CDaR under the simulated scenarios. The VaR with significant level $\eta$ under the simulated scenario is estimated by
\begin{align}
\textup{VaR}_\eta(R_m(x)) = -\inf\left\{u\,\big|\,\frac{1}{S}\sum_{s=1}^S 1_{R_m^s(x)<u}> 1-\eta\right\}.
      \label{eq:VaR scenario}
\end{align}

Let $R_m^{(k)}(x)$ be the $k$-th smallest value in $\{R_m^s(x)\,|\, s=1,2,\cdots, S\}$, then CVaR at the significant level $\eta$ is estimated according to the formula
    \begin{align}
\nonumber &      \CVaR_{\eta}(R_m(x)) 
      \\&= -\frac{1}{\eta}\LR{\frac{1}{S}\sum_{k=1}^{\lceil S(\eta)\rceil-1}R_m^{(k)}(x)+\LR{\eta-\frac{\lceil S\eta \rceil-1}{S}}R_m^{(\lceil S\eta\rceil)}(x)},
      \label{eq:CVaR scenario}
    \end{align}
where $\lceil x\rceil$ denotes the smallest integer larger than $x$ (See \citeauthor{StoyanovRachevFabozzi:2009} \citeyear{StoyanovRachevFabozzi:2009} for details).

The rate of return of the $n$-th asset between time $t_{i-1}$ and time $t_{i}$ is defined by
\[
r_n(t_i, \omega) = \frac{P_n(t_i,\omega)-P_n(t_{i-1},\omega)}{P_n(t_{i-1},\omega)}, ~~~ t>0,
\]
with $r_n(0, \cdot)=0$. It is referred to as return in this paper.
Let $x=(x_1, x_2, \cdots, x_N)^T\in\R^N$ be a capital allocation vector of a long-only portfolio satisfying $\sum_{n=1}^N x_n = 1$ and $x_n\in[0,1]$ for all $n\in\{1,2,\cdots, N\}$. Then, the portfolio return $U(x, t_i,\omega_s)$ between time $t_{i-1}$ and time $t_i$ is 
\[
U(x, t_i, \omega_s) = \sum_{n=1}^N x_n r_n(t_i,\omega_s).
\]

The scenario-based estimation on  uncompounded cumulative return at time $t_i$ is 
\[
Q(x, t_i, \omega_s) = \sum_{t=1}^t   U_n(x,t_i,\omega_s).
\]

Denote $Q(x, t_i, \omega_s)$ by $Q_i^s(x) $. Using \eqref{eq:DD}--
\eqref{eq:MDD},  DD, ADD, and MDD on a single simulated scenario are estimated, respectively, by
\begin{align} \label{DDdef}
\textup{DD}_{m,s} &:= \textup{DD}(t_m,\omega_s) = \max_{j\in\{0,1,2,\cdots, m\}} Q_j^s(x) - Q_m^s(x),\\
\textup{ADD}_{M,s} &:= \textup{ADD}(t_M, \omega_s) = \frac{1}{M}\sum_{m=1}^M \textup{DD}_{m,s},\\
 \textup{MDD}_{M,s} &:= \textup{MDD}(t_M, \omega_s) = \max_{m\in\{1,2,\cdots, M\}} \textup{DD}_{m,s}.
\end{align}

Furthermore, using \eqref{eq:FDD}--
\eqref{eq:CDaR}, CDaR is estimated on a single simulated scenario by
\[
\textup{CDaR}_\eta(T, s) := \textup{CDaR}_\eta(T, \omega_s) = \frac{1}{(1-\eta)M}\sum_{m=1}^M \textup{DD}_{m,s} 1_{\textup{DD}_{m,s}>\zeta_\eta^s},
\]
where
\[
\zeta_\eta^s := \zeta_\eta(w_s) = \begin{cases}
\inf\{ z|F_{\textup{DD}}(z,s)\ge\eta\} &\text{ if } \eta\in(0,1]\\
0 &\text{ if } \eta=0
\end{cases}.
\]
with
\[
F_{\textup{DD}}(z,s) = \frac{1}{M}\sum_{m=1}^M 1_{\textup{DD}_{m,s}\le z}.
\]

Equivalently, it is estimated by the optimization formula
\[
\textup{CDaR}_\eta(T, s)  = \min_{\zeta}\left\{\zeta+\frac{1}{(1-\eta)M}\sum_{m=1}^M \max\left(\textup{DD}_{m,s}-\zeta, 0\right)\right\}.
\]

With \eqref{eq:FDD_multi}--
\eqref{eq:CDaR_multi}, DD and CDaR on multiple scenarios are estimated by
\[
\textup{DD}(m) = \frac{1}{S}\sum_{s=1}^S \textup{DD}_{m,s}
\]
and
\begin{align}
\nonumber &\textup{CDaR}_\eta(T) \\
&= \left(\frac{F_{\textup{DD}}(\zeta(\eta))-\eta}{1-\eta}\right)\zeta(\eta)+\frac{1}{(1-\eta) M}\sum_{s=1}^S \sum_{m=1}^M \frac{1}{S}\textup{DD}_{m,s}1_{\textup{DD}_{m,s}>\zeta(\eta)},
\label{eq:CDaR_multi_scenario}
\end{align}
respectively, where
\[
F_{\textup{DD}}(z) = \frac{1}{S}\sum_{s=1}^S F_{\textup{DD}}(z,s)
\]
and
\[
\zeta_\eta = \begin{cases}
\inf\left\{ z\big|F_{\textup{DD}}(z)\ge\eta\right\} &\text{ if } \eta\in(0,1]\\
0 &\text{ if } \eta=0
\end{cases}.
\]

In this paper, we will apply time series models to generate the scenario of $r_n(m,s)$ for $m\in\{1,2,\cdots, M\}$, $s\in\{1,2,\cdots, S\}$ and $n\in\{1,2,\cdots, N\}$.

\subsection{Multivariate Normal Tempered Stable Distribution}

Let $\Tau$ be a strictly positive random variable defined by the characteristic function for $\lambda\in(0,2)$ and $\theta>0$
\begin{equation*}\label{NTScharac}
    \phi_\Tau(u) = \exp\left(-\frac{2\theta^{1-\frac{\lambda}{2}}}{\lambda}((\theta-iu)^{\frac{\lambda}{2}}-\theta^{\frac{\lambda}{2}})\right).
\end{equation*}

Then, $\Tau$ is referred to as a tempered stable subordinator with parameters $\lambda$ and $\theta$.
Suppose that $\bm{\xi}=(\xi_1, \cdots, \xi_N)^T$ is a $N$-dimensional standard normal distributed random vector with a $N\times N$ covariance matrix $\Sigma$, i.e., $\bm{\xi}\sim \Phi(0,\Sigma)$.
The $N$-dimensional NTS distributed random vector $X = (X_1,...,X_N)^T$ is defined by
\begin{equation*}\label{MNTS}
\bm{X} = \bm{\mu} + \bm{\nu}(T-1)+\sqrt{T}\diag(\bm{\gamma})\bm{\xi},
\end{equation*}
where $\bm{\mu}=(\mu_1, \cdots, \mu_N)^T \in \mathbb{R}^N$, $\bm{\nu}=(\nu_1, \cdots, \nu_N)^T \in \mathbb{R}^N$, $\bm{\gamma}=(\gamma_1, \cdots, \gamma_N)^T \in \mathbb{R}_{+}^N$ and $\Tau$ is independent of $\bm{\xi}$. The $N$-dimensional multivariate NTS distribution specified above is denoted by
\[
X\sim\textup{MNTS}_N(\lambda, \theta, \bm{\nu}, \bm{\gamma}, \bm{\mu},\Sigma).
\] 

Let $\mu_n = 0$, and $\gamma_n = \sqrt{1-\nu_n^2 (\frac{2-\lambda}{2\theta})}$, $ \lvert \nu_n \rvert < \sqrt{\frac{2\theta}{2-\lambda}}$ for $n\in\{1,\cdots, N\}$. This yields a random variable having zero mean and unit variance. In this case, the random variable is referred to as \emph{standard NTS} distributed random variable and denoted by $X\sim\textup{stdMNTS}_N(\lambda, \theta, \bm{\nu}, \Sigma)$. The covariance matrix of $X$ is given by
\begin{equation*}\label{MNTS}
\Sigma_X = \text{diag}(\bm{\gamma}) \Sigma \text{diag}(\bm{\gamma}) + \frac{2-\lambda}{2\theta} \bm{\nu}^\top \bm{\nu}.
\end{equation*}

\subsection{Regime-Switching GARCH Model}
GARCH($p,q$) model has been studied extensively as a model for volatility. 
\begin{equation*}
\begin{aligned}
r_t &= \eta + \sigma_t\epsilon_t,\\
u_t &= \sigma_t\epsilon_t,\\
\sigma_t^2 &= \omega + \sum	_{i=1}^p \alpha_i u_{t-i}^2 + \sum_{i=1}^q \beta_i\sigma_{t-i}^2.
\end{aligned}
\end{equation*}
where $r_t$ is the return at time $t$, $\sigma_t$ is the variance at time $t$, $\eta,\omega,\alpha_i,\beta_i$ are parameters. We will use GARCH(1,1) in this paper.

Before specifying our model, it would be clear to first specify the univariate Markov regime-switching model in \cite{Haas1} {as follows} 
\begin{equation}\label{mrsgarch}
\begin{aligned}
r_t &= \eta_{\Delta_t}+\sigma_{\Delta_t,t}\epsilon_t,\\
u_t&= \sigma_{\Delta_t,t}\epsilon_t,\\
{\bm{\sigma}_t^2} &= \bm{\omega} + \bm{\alpha} u_{t-1}^2 + \bm{\beta}\circ{\bm{\sigma}_{t-1}^2},\\
\epsilon _{t}& \overset{\text{iid}}\sim \text{N}(0,1) ,
\end{aligned}
\end{equation}
where 
\begin{itemize}
\item $k=$ number of regimes in a Markov chain;
\item $\Delta_t=\{1,\cdots,k\}=$ Markov chain with an irreducible and primitive transition matrix that determines the regime at time $t$;
\item $r_t=$  return at time $t$;
\item $\bm{\sigma}_t^2 = (\sigma_{1,t}^2,....,\sigma_{k,t}^2)=$ vector of parallel variance;
\item  $\bm{\omega} = (\omega_1,....,\omega_k)$, $\bm{\alpha} = (\alpha_1,....,\alpha_k)$, $ \bm{\beta} = (\beta_1,....,\beta_k) $. $\bm{\omega}$, $\bm{\alpha}$, $\bm{\beta}$ are vectors of coefficients;
\item $\eta_{\Delta_t}=$  regime specific mean return in regime $\Delta_t$ at time $t$;
\item $\sigma_{\Delta_t,t} = $  the standard deviation in regime $\Delta_t$ at time $t$;
\item $\circ$ denotes  element-wise product.
\end{itemize}

There are $k$ parallel GARCH models evolving simultaneously. The Markov chain $\Delta_t$ determines which GARCH model is realized at next moment.

To better understand (\ref{mrsgarch}), let us create the following matrix
\[
[\sigma_{\Delta_t,t}]_{\Delta_t = 1,\ldots,k, \; t=1,\ldots T} = 
	\begin{bmatrix} 
    \sigma_{1,1} & \sigma_{1,2} &\dots & \sigma_{1,t} & \dots \\
    \sigma_{2,1} & \sigma_{2,2} &\dots & \sigma_{1,t} & \dots \\
    \vdots    		& \vdots 			& 	\ddots 				&\dots			 &  \dots		\\
    \sigma_{k,1} & \sigma_{k,2} &\dots   & \sigma_{k,t} & \dots
    \end{bmatrix} .
\]

Each column is a total of $k$ parallel standard deviations at time $t$. Each row is the process of one regime through time 1 to $T$. The process of standard deviation is generated as follows:

\noindent Step 1: At $t=1$, the initial column is generated.

\noindent Step 2: $\Delta_1$ decides which row is realized.

For example, suppose $\Delta_1 = 2$. That is, $\sigma_{2,1}$ is realized in real-world at $t=1$.

\noindent Step 3: By vector calculation, 2nd column is generated.

For example, suppose $\sigma_{2,1}$ is realized at $t=1$. We calculate
\[
\begin{bmatrix} 
    \sigma_{1,2} \\
    \sigma_{2,2} \\
    \vdots  		\\
    \sigma_{k,2} 
    \end{bmatrix}
=
\begin{bmatrix} 
    \omega_1 \\
    \omega_2 \\
    \vdots  		\\
    \omega_k 
    \end{bmatrix}
+
\sigma_{2,1}\epsilon_1
\begin{bmatrix} 
    \alpha_1 \\
    \alpha_2 \\
    \vdots  		\\
    \alpha_k
    \end{bmatrix}
+
\begin{bmatrix} 
    \beta_1 \\
    \beta_2 \\
    \vdots  		\\
    \beta_k 
    \end{bmatrix}
\circ
\begin{bmatrix} 
    \sigma_{1,1} \\
    \sigma_{2,1} \\
    \vdots  		\\
    \sigma_{k,1} 
    \end{bmatrix} .
\]

\noindent Step 4: $\Delta_2$ Decides which row is realized.

For example, suppose $\Delta_2 = 1$. That is, $\sigma_{1,2}$ is realized in real world at $t=2$. A regime switch from regime $2$ to regime $1$ takes place.

\noindent Step 5: Repeat Step 1 to Step 4.

It is important to note that in the example, the realized variance at time $t=2$, $\sigma_{1,2}$, is calculated with both the realized variance at time $t=1$, $\sigma_{2,1}$, and unrealized $\sigma_{1,1}$. This is obvious from the vector equation shown above. $\sigma_{1,1}$ is in the parallel process and is not realized in real world. 

More generally, when there is a regime switch, from the vector equation ${\bm{\sigma}_t^2} = \bm{\omega} + \bm{\alpha} {u_{t}}^2 + \bm{\beta}\circ{\bm{\sigma}_{t-1}^2}$, we know that the variance ${\sigma}_{\Delta_t,t}$ is determined by ${\sigma}_{\Delta_t,t-1}$ rather than ${\sigma}_{\Delta_{t-1},t-1}$. ${\sigma}_{\Delta_t,t-1}$ is a variance in regime $\Delta_t$ at time $t$, which is generated simultaneously with ${\sigma}_{\Delta_{t-1},t-1}$, but does not exist in reality at time $t$ or $t-1$. This makes the model different from the one in \cite{Henneke} that requires all parameters to switch regime according to a Markov chain. In that model, there is no parallel process, and ${\sigma}_{\Delta_t,t}$ is determined by ${\sigma}_{\Delta_{t-1},t-1}$.

{The stationary conditions derived in \cite{Haas1} require definitions of the~matrices: }
\begin{equation*}\label{stationary}
M_{ji} = p_{ji}(\bm{\beta}+\alpha e_{i}^\top), \quad i,j = 1,...,k
\end{equation*}
and the block matrix
\begin{equation*}\label{stationary2}
M= [M_{ji}], \quad i,j =1,....,k.
\end{equation*}

The necessary and sufficient condition for stationarity is $\rho(M)<1$, where $\rho	(M)$ denotes the largest eigenvalue of matrix $M$. 

\section{MRS-MNTS-GARCH Model}\label{sec_Model}

This section defines the MRS-MNTS-GARCH model and introduces the procedures for model fitting and sample path simulation.

\subsection{Model Specification}

To model a multivariate process, we assume that the variance of each individual asset evolves independently according to the univariate regime-switching GARCH model specified previously with possibly different number of regimes, while the correlation between assets is modeled separately by stdMNTS distribution in the joint standard innovations. The joint innovations is defined bya HMM with stdMNTS as conditional distribution in each regime.

The MRS-MNTS-GARCH model is defined by
\begin{equation*}\label{Modelspec}
\begin{aligned}
\bm{r}_t &= \bm{\eta}_{t}+\bm{\sigma}_{t}      \circ \bm{\epsilon}_t\\
\bm{u}_t&= \bm{\sigma}_{t}      \circ \bm{\epsilon}_t \\
\left({\bm{\sigma}_t^{(i)}}\right)^2 &= \boldsymbol{\omega}^{(i)} + \boldsymbol{\alpha}^{(i)} {\left(u_{t-1}^{(i)} \right)}^2 + \boldsymbol{\beta}^{(i)}\circ    \left({\bm{\sigma}_{t-1}^{(i)}}\right)^2,i=1,\ldots,N\\
\bm{\epsilon} _{t}& \overset{\text{iid}}\sim \text{stdMNTS}(\lambda_{\Delta_t^{D}},\theta_{\Delta_t^{D}},\bm{\nu}_{\Delta_t^{D}},\Sigma_{\Delta_t^{D}}),
\end{aligned}
\end{equation*}
where 
\begin{itemize}
\item $N=$ number of assets;
\item  $k^{(i)} = $   number of regimes of the $i$-th asset;
\item  $\Delta^{(i)} = \{ 1,\cdots, k^{(i)}  \} =$ Markov chain with an irreducible and primitive transition matrix that determines the regime of the $i$-th asset at time $t$;
\item $\Delta_t^{(i)} = \{ 1,\cdots, k^{(i)}  \} =$ realization of $\Delta^{(i)}$ at time $t$;
\item  $\Delta^{D} = \{ 1,\cdots, k^{D}  \} =$ Markov chain with an irreducible and primitive transition matrix that determines the regime of the joint innovations at time $t$;
\item $\Delta_t^D = \{ 1,\cdots, k^D  \} =$ realization of $\Delta^D$ at time $t$;
\item $\bm{r}_t = (r_t^{(1)},\ldots,r_t^{(N)}) =$ vector return of $N$ assets at time $t$;
\item $\bm{\eta}_t = (\eta_{\Delta_t^{(1)}}^{(1)},\ldots,\eta_{\Delta_t^{(N)}}^{(N)}) =$ vector mean return at time $t$;
\item $\eta_{\Delta_t^{(i)}}^{(i)}=$  mean return of the $i$-th asset at time $t$, $\eta_{\Delta_t^{(i)}}^{(i)}$ is determined by the regime of the $i$-th asset at time $t$, $\Delta_t^{(i)}$;
\item $\bm{\sigma}_{t} = \left(\sigma_{\Delta_t^{(1)},t}^{(1)},\ldots,\sigma_{\Delta_t^{(N)},t}^{(N)} \right)=$  vector of standard deviation of $N$ assets at time $t$;
\item${\bm{\sigma}_t^{(i)}}= \left( {    \sigma_{1,t}^{(i)}    },\ldots,{ \sigma_{k^{(i)},t}^{(i)}  }  \right)=$  standard deviation vector of the $i$-th asset with $\left({\bm{\sigma}_t^{(i)}}\right)^2 = \left( {  \left(  \sigma_{1,t}^{(i)}  \right)  }^2,\ldots,{\left(  \sigma_{k^{(i)},t}^{(i)} \right)  }^2  \right)$;
\item $u_{t-1}^{(i)} =$ the $i$-th element of $\bm{u}_t$, with $\bm{u}_t = \left( u_t^{(1)},\ldots,u_t^{(N)} \right)$;
\item $\bm{\omega}^{(i)} = (\omega_1^{(i)},\ldots,\omega_{k^{(i)}}^{(i)})$, $\bm{\alpha}^{(i)} = (\alpha_1^{(i)},\ldots,\alpha_{k^{(i)}}^{(i)})$, $\bm{\beta}^{(i)} = (\beta_1^{(i)},\ldots,\beta_{k^{(i)}}^{(i)})$. $\bm{\omega}^{(i)}$, $\bm{\alpha}^{(i)}$, $\bm{\beta}^{(i)}$ are the vector of coefficients of the $i$-th asset in $k^{(i)}$ regimes;
\item $\bm{\epsilon}_t = (\epsilon_t^{(1)},\ldots,\epsilon_t^{(N)}) =$ joint innovations;
\item  $\circ$ denotes element-wise product.
\end{itemize}

For each asset, the volatility process is the same as that of the univariate regime-switching GARCH model in (\ref{mrsgarch}). It is a classical GARCH process when it does not shift within regimes. Unlike the most flexible model in which the variance at time $t$ is calculated with the variance at time $t-1$ in the last regime, when a regime shift takes place at time $t$, the variance in the model at time $t$ is determined by the variance at time $t-1$ within the new regime.  The regimes of each asset and joint innovations evolve independently, mitigating the difficulty in estimation caused by all parameters switching regimes simultaneously.

\subsection{Parameter Estimation}

Although NTS distribution is flexible enough to accommodate many stylized facts of asset return, the absence of an analytical density function presents some difficulties in estimation. Our estimation methodology is adapted from \citeauthor{Kim1} (\citeyear{Kim1}, \citeyear{Kim2}), 
which are among the first works to incorporate NTS distribution in portfolio optimization. Another estimation of stdMNTS distribution with EM algorithm combined with fast Fourier transform is studied in \cite{Bianchi}. Below we describe the procedure to fit the~model.

First, we fit the univariate Markov regime-switching GARCH model with $t$ innovation on the index return and extract the innovations and the regimes at each time $t$, $\Delta_t^D$. 
It is intuitive to assume that the correlation between assets change drastically when a market regime shift takes place. Thus, here we assume  that the Markov chain $\Delta^D$ that determines the regime of innovations  is the same as the Markov chain that determines the regime of the market index. That is, the sample path of the regime of the innovation is derived by extracting the sample path of the regime of the market index. The innovations of the  model calibrated by the market index are used to estimate tail parameters $\lambda_{\Delta^D}, \theta_{\Delta^D}$ in each regime. 

Next, we fit the univariate model on each asset and extract the innovations for subsequent estimation of stdMNTS distribution. In each regime, the stdMNTS distribution has two tail parameters $\lambda_{\Delta^D}, \theta_{\Delta^D}$ and one skewness parameter vector $\bm{\nu}_{\Delta^D}$. Common tail parameters $\lambda_{\Delta^D}, \theta_{\Delta^D}$ are assumed for individual constituents, which are estimated from the market index in each regime. The skewness parameter $\bm{\nu}_{\Delta^D}$ is calibrated by curve fitting for each asset. 

Finally,  we plug the empirical  covariance matrix of the innovations in each regime ${\Sigma_X}_{\Delta^D}$ in   the formula $\Sigma_{\Delta^D} = \text{diag}(\bm{\gamma}_{\Delta^D})^{-1}({\Sigma_X}_{\Delta^D} -  \frac{2-\lambda_{\Delta^D}}{2\theta_{\Delta^D}} \bm{\nu}_{\Delta^D}^\top \bm{\nu}_{\Delta^D})\text{diag}(\bm{\gamma}_{\Delta^D})^{-1}$ to compute $\Sigma_{\Delta^D}$. Note that $\Sigma_{\Delta^D}$ is supposed to be a positive definite matrix. To guarantee this, we can either substitute $\bm{\nu}_{\Delta^D}$ with the closest vector to $\bm{\nu}_{\Delta^D}$ in $L^2$ norm which renders ${\Sigma_X}_{\Delta^D} -  \frac{2-\lambda_{\Delta^D}}{2\theta_{\Delta^D}} \bm{\nu}_{\Delta^D}^\top \bm{\nu}_{\Delta^D}$ positive definite matrix, or directly find the closest positive definite matrix to $\Sigma_{\Delta^D}$. We choose the second method. Furthermore, the usual issue of estimating covariance matrix arises here as well, especially when a regime only has a short lasting period. To address this issue, we apply the denoising technique in \cite{Laloux} on the  positive definite matrix derived earlier to estimate $\Sigma_{\Delta^D}$.

The number of regimes of each asset is set as equal to that of the market index. We find that univariate Markov switching model with more than three regimes often has one unstable regime that lasts for a very short period and switches frequently and sometimes has one regime with clearly bimodal innovations distribution. Thus, it is desirable to limit the number of regimes smaller than 4, check unimodal distribution with Dip test and choose the one with the highest BIC value. 

The procedure is summarized as follows.

\noindent Step 1: Fit the univariate Markov regime-switching GARCH model with t innovation on the index return and extract the innovations and the sample path of regimes.

\noindent Step 2: Estimate common tail parameters $\lambda_{\Delta^D}, \theta_{\Delta^D}$ in each regime.

\noindent Step 3: Fit univariate model on each asset and extract the innovations.

\noindent Step 4: Estimate skewness parameters $\nu_{\Delta^D}$ for assets in each regime, calculate  $ \gamma_{\Delta^D} = \sqrt{1-{\nu_{\Delta^D}}^2 (\frac{2-\lambda_{\Delta^D}}{2\theta_{\Delta^D}})}$.

\noindent 
{ Step 5}: Calculate the matrix 
$\text{diag}(\bm{\gamma}_{\Delta^D})^{-1}({\Sigma_X}_{\Delta^D} -  \frac{2-\lambda_{\Delta^D}}{2\theta_{\Delta^D}} \bm{\nu}_{\Delta^D}^\top \bm{\nu}_{\Delta^D})\text{diag}(\bm{\gamma}_{\Delta^D})^{-1}$, find the closest positive definite matrix and apply the denoising technique to estimate $\Sigma_{\Delta^D}$.

\subsection{Simulation}
To conduct simulation with the calibrated model,
we follow the procedures specified~below.

\noindent Step 1: Simulate $S$ sample paths of standard deviation of $N$ assets for $T$ days with the fitted~model.

\noindent Step 2: Simulate $S$ sample paths of a Markov chain $\Delta^D$ with the market's transition matrix. Calculate the total number of each regime in all sample paths.

\noindent Step 3: Draw i.i.d. samples from stdMNTS distribution in each regime. The number of samples is determined by the calculation in  Step 2. To draw one sample, we follow the procedure below.

\noindent Step 3.1: Sample from multivariate normal distribution N(0,$\Sigma_{\Delta^D}$).

\noindent Step 3.2: Sample from the tempered stable subordinator $T$ with parameters $\lambda_{\Delta^D}$ and $\theta_{\Delta^D}$.

\noindent Step 3.3: Calculate $\bm{X} =  \bm{\nu}_{\Delta^D}(T-1)+\sqrt{T}(\bm{\gamma}_{\Delta^D} \circ \bm{\xi}_{\Delta^D})$.

\noindent Step 4: Multiply the standard deviation in  Step 1 with standard joint innovations in  Step 3 accordingly. Add regime-specific mean $\bm{\eta}_{\Delta^D}$ to get simulated asset returns.

\section{Portfolio Optimization} \label{sec_Portfolio_Optimization}

In this section, we discuss issues on portfolio optimization.
\subsection{Problem Formulation}
Conditioned on given sample path(s), classical portfolio optimization is formulated as a trade-off between risk and return   
\begin{equation*}\label{riskreturn}
\begin{aligned}
 \min_{{x}}\: &  W(\Rc(x, t, \omega)|t\in[0,T], \; \omega \in \Omega ) \   \\    
  \text{s.t.}\:&  E \left[ \Rc(x, T, \omega) | \omega \in \Omega  \right]   \geq d,\\
    &{x} \in V,
\end{aligned}\\
\end{equation*}
where $d$ is the benchmark return, $\Rc(x, t, \omega)$ is the portfolio return at time $t\in[0,T]$ in given sample paths $\omega\in\Omega$, $W(\Rc(x,t, \omega)|t\in[0,T])$ is a risk measure of $\Rc(x, t, \omega)$ for time $t\in[0,T]$ estimated based on given sample paths $\Omega$, and ${x} = (x_1,\ldots,x_n)$ is the allocation vector of $n$ assets.  A typical setting of $V$ is $\{{x} | \sum_{i=1}^N x_i =1, x_i \geq 0 \}$, meaning that short selling is not allowed.

\subsection{Discussion on Risk Measures and Constraint}

If the historical sample paths of asset returns are used as input to the CVaR  optimization,  we  minimize the CVaR  of daily portfolio return from time $t=1$ to $t=T$. In the setting where we use simulation as input, we can choose to minimize the CVaR of daily returns from time $t=1$ to $t=T$ or CVaR of compounded returns at time $t=T$. We can set constraints on average daily returns from time $t=1$ to $t=T$ or compounded portfolio  return at time $t=T$.  Notice that the daily return in the simulation of GARCH model is not i.i.d. Furthermore, it is reasonable to focus on the the return  at the final time. Thus, we  choose to minimize the CVaR of return at time $t=T$ instead of return from time $t=1$ to $t=T$ subject to constraint on compounded portfolio  return at time $t=T$.

The definition of drawdown needs some clarification as well. 
The portfolio drawdown at time $t$ is defined in (\ref{DDdef}).
\[
\textup{DD}_{m,s} : = \max_{j\in\{0,1,2,\cdots, m\}} Q(x, t_j, \omega_s) - Q(x, t_m, \omega_s) = \sum_{i=h}^m  r(x,t_i,w_s),
\]
where $h=\arg\max_{j\in\{0,1,2,\cdots, m\}} Q(x, t_j, \omega_s)$, 
is the sum of daily returns from its peak. It is somewhat unconventional at first sight compared with the absolute drawdown defined by
\[
\textup{absDD}_{m,s} : = \max_{j\in\{0,1,2,\cdots, m\}} P(x, t_j, \omega_s) - P(x, t_m, \omega_s) = P(x, t_h, \omega_s) \prod_{i=h}^m (1+ r(x,t_i,w_s))
\]
and the relative drawdown defined by
\[
\textup{reDD}_{m,s} : = \frac{\max_{j\in\{0,1,2,\cdots, m\}} P(x, t_j, \omega_s) - P(x, t_j,\omega_s)}{\max_{j\in\{0,1,2,\cdots, m\}} P(x, t_j, \omega_s)} = \prod_{i=h}^m (1+ r(x,t_i,w_s)),
\]
where $h=\arg\max_{j\in\{0,1,2,\cdots, m\}} Q(x, t_j, \omega_s)$.
CDaR is essentially an average of drawdowns that exceed the threshold. CDaR defined in (\ref{eq:CDaR_multi_scenario}) with $\text{DD}_{m,s}$ satisfies good properties such as nonnegativity, insensitivity
to constant shift, positive homogeneity and convexity (See \citeauthor{Uryasev1} \citeyear{Uryasev1}), which gives great advantage in developing optimization techniques. 
CDaR can be defined in a similar way with $\textup{absDD}_{m,s}$ and $\textup{reDD}_{m,s}$.

It is possible to use the algorithm designed for CDaR defined by $\textup{DD}_{m,s}$ to minimize CDaR defined by $\textup{absDD}_{m,s}$. By substituting the input $r(x,t_i,w_s)$ with $P(x, t_i, \omega_s) - P(x, t_{i-1},\omega_s)$, the calculation of $\textup{DD}_{m,s}$ becomes
\[
\sum_{i=h}^m P(x, t_i, \omega_s) - P(x, t_{i-1},\omega_s) = \textup{absDD}_{m,s} 
\]
where $h=\arg\max_{j\in\{0,1,2,\cdots, m\}} Q(x, t_j, \omega_s)$.
However, it is obvious from the definitions  that CDaR defined $\textup{absDD}_{m,s}$ and $\textup{reDD}_{m,s}$ do not satisfy basic axioms such as positive homogeneity. 

Another drawback of absolute drawdown is that when we consider a long time period, drawdowns could vary greatly in absolute value but be close in relative value, the latter of which we are more concerned about. Uncompounded cumulative return is determined by rate of returns regardless of the price. This discussion justifies our definition of drawdown with $\text{DD}_{m,s}$.

\section{Empirical Study} \label{sec_Empirical_Study}
In this section, we demonstrate the superiority of our approach with various in-sample and out-of-sample tests. We conduct model fitting, sample path simulation and portfolio optimization. In the in-sample test, we conduct a KS test on the marginal NTS distribution and marginal $t$-distribution to show that the NTS distribution has a much better fit on the innovations. We also provide the transition matrix and the correlation matrices of the joint innovations in each regime. In the out-of-sample test, we use various performance measures and visualizations for the rolling-window studies in recent years to show the outperformance of our portfolio.

For diversification purposes, we set the range of weights as $[0.01, 0.15]$. A case study in \cite{Uryasev3} found that without constraints on weights, the optimal portfolio only consists of a few assets among hundreds.  
{Investors often empirically categorize the market as bull, bear and sideways market. \cite{Haas1} experiments with two and three regimes.} We limit the number of regimes to be smaller than 
or equal to
three. The Dip test is conducted on innovations to ensure unimodal innovation distribution. Starting fitting with 
three regimes, a model with fewer regimes is used instead when the $p$-value is lower than $0.1$ or have a regime lasting shorter than $40$ days in total.  Preliminary tests show that   without these selection criteria, the model would have a risk of overfit in some cases.  For example, the innovation could be multimodal, and the regime switches hundreds of times within the tested period.  
{The phenomena is even more pronounced when the number of regimes is higher than three.} To be concise in notation, we denote the portfolios in a convenient manner. For example, 0.9-CDaR portfolio denotes the optimal portfolio derived from the optimization with 0.9-CDaR as the risk measure. The standard deviation optimal portfolio is the optimal portfolio in the classical Markowitz model that maximizes Sharpe ratio, which we also denoted by mean-variance (MV) optimal portfolio. The word optimal is sometimes omitted. Since we simulate returns for $10$ days, all CVaR with varying confidence levels are calculated with the simulated cumulative returns on the $10$-th day. We omit the time and denote the optimal portfolio simply as CVaR portfolio.

\subsection{Data}
The data are comprised of the adjusted daily price of DJIA index, 29 of the constituents of DJIA Index and three ETFs of bond, gold and shorting index (Tickers: TMF, SDOW and UGL) from January 2010 to September 2020. One constituent DOW is removed because it is not included in the index throughout the time period. Since we use 1764 trading days' (about 7 years) data to fit the model, the time period for out-of-sample rolling-window test is from January 2017 to September 2020, which includes some of the most volatile periods in history caused by the a pandemic and some trade tensions. 
By reweighting the index constituents, it can be regarded as an enhanced index portfolio.

\subsection{Performance Measures}
{Various performance measures are discussed in \cite{Biglova}, \mbox{\cite{StoyanovRachevFabozzi:2009}}, \cite{Cheridito}.
Similar to Markowitz Model, an efficient frontier can be derived by changing the value of constraint. Each portfolio on the efficient frontier is optimal in the sense that no portfolio has a higher return with the same value of risk measure. In this paper, we refer to the portfolio with the highest performance ratio $\frac{E  \left[ R(x,T,\cdot)  \right] }{W( R(x,t,\cdot)|t\in[0,T])}$ as an optimal portfolio and the others on the efficient frontier as suboptimal portfolios.  Note that $E  \left[ R(x,T,\cdot)  \right] $ is $\text{CVaR}_0( R(x,T,\cdot) )$. Thus, the mean-CVaR ratio can be regarded as a special case of the Rachev ratio $\frac{\text{CVaR}_{\eta_1}( R(x,T,\cdot) )}{\text{CVaR}_{\eta_2}( R(x,T,\cdot) )}$. We test with $\eta_1=\eta_2=0.1$ to focus on the distribution~tails.}

\subsection{In-Sample Tests}
We present various in-sample test results. The time period during which 2-regime model is first selected is used in fitting is from 30 April 2010 to 5 May  2017. $100$ days is classified as regime $1$, and $1665$ days is classified as regime $2$. The standard deviation of the DJIA index is $0.00217$ in regime $1$ and $0.00801$ in regime $2$. Regime $2$ is more volatile than regime $1$. One might expect that the market is in a less volatile regime longer and has a spiked standard deviation during a short but much more volatile regime. It is not the case in this tested period according to our model.

\subsubsection{Kolmogorov–Smirnov Tests on Marginal Distributions}

We conduct a Kolmogorov–Smirnov test (KS test) on the marginal distributions of the joint innovation of MRS-MNTS-GARCH model in the tested period. The results are presented in Table \ref{KS}.
For comparison, we also fit multivariate $t$-distribution on the joint innovations and report the results of KS tests on the marginal distributions. The degree of freedom of multivariate $\textit{t}$-distribution in regime $1$ and regime $2$ are $5.34$ and $5.68$, respectively. Note that the marginal distribution of stdMNTS distribution is still NTS distribution. 
Since we assume common parameters $\lambda$ and $\theta$ in MNTS distribution, we report the parameter $\beta$ of each asset in both regimes. The data are rounded to $3$ decimals~place.

\begin{table}[!htbp]
  \caption{Kolmogorov–Smirnov tests on marginal distributions of joint innovation} \label{KS}
\begin{threeparttable}\small \centering
 
\renewcommand{\arraystretch}{0.8}
\begin{tabular}{@{\extracolsep{5pt}} ccccccc} 
\\\hline 
\\[-1.8ex] 
& Regime & beta & KS Statistics of NTS & $p$-value & KS Statistics of t & $p$-value \\ 
\hline \\[-1.8ex] 
AAPL & 1 & 2.048 & 0.129 & 0.067 & 0.125 & 0.081 \\ 
 & 2 & 0.038 & 0.018 & 0.624 & 0.055 & 0 \\ 
AMGN & 1 & 1.133 & 0.054 & 0.917 & 0.058 & 0.875 \\ 
 & 2 & 0.139 & 0.012 & 0.955 & 0.045 & 0.002 \\ 
AXP & 1 & -0.513 & 0.022 & 1 & 0.027 & 1 \\ 
 & 2 & -0.021 & 0.022 & 0.392 & 0.06 & 0 \\ 
BA & 1 & -1.129 & 0.016 & 1 & 0.039 & 0.996 \\ 
 & 2 & -0.075 & 0.009 & 0.999 & 0.041 & 0.007 \\ 
CAT & 1 & 1.681 & 0.083 & 0.476 & 0.09 & 0.369 \\ 
 & 2 & 0.069 & 0.012 & 0.973 & 0.043 & 0.004 \\ 
CRM & 1 & -0.861 & 0.041 & 0.993 & 0.045 & 0.982 \\ 
 & 2 & 0.095 & 0.032 & 0.071 & 0.071 & 0 \\ 
CSCO & 1 & 1.297 & 0.084 & 0.454 & 0.066 & 0.75 \\ 
 & 2 & 0.058 & 0.043 & 0.004 & 0.08 & 0 \\ 
CVX & 1 & -1.523 & 0.064 & 0.781 & 0.063 & 0.791 \\ 
 & 2 & -0.055 & 0.019 & 0.563 & 0.029 & 0.11 \\ 
DIS & 1 & -0.644 & 0.115 & 0.132 & 0.103 & 0.225 \\ 
 & 2 & 0.025 & 0.014 & 0.878 & 0.053 & 0 \\ 
GS & 1 & 1.274 & 0.038 & 0.998 & 0.031 & 1 \\ 
 & 2 & -0.046 & 0.014 & 0.917 & 0.036 & 0.026 \\ 
HD & 1 & -1.431 & 0.035 & 0.999 & 0.056 & 0.901 \\ 
 & 2 & 0.025 & 0.01 & 0.995 & 0.047 & 0.001 \\ 
HON & 1 & -1.496 & 0.064 & 0.783 & 0.064 & 0.776 \\ 
 & 2 & 0.131 & 0.015 & 0.843 & 0.054 & 0 \\ 
IBM & 1 & 0.64 & 0.023 & 1 & 0.041 & 0.993 \\ 
 & 2 & -0.019 & 0.019 & 0.587 & 0.057 & 0 \\ 
INTC & 1 & -1.551 & 0.045 & 0.981 & 0.085 & 0.443 \\ 
 & 2 & 0.061 & 0.008 & 1 & 0.044 & 0.003 \\ 
JNJ & 1 & 1.87 & 0.055 & 0.908 & 0.109 & 0.174 \\ 
 & 2 & -0.013 & 0.008 & 1 & 0.041 & 0.007 \\ 
JPM & 1 & 1.857 & 0.023 & 1 & 0.083 & 0.464 \\ 
 & 2 & 0.026 & 0.008 & 1 & 0.041 & 0.007 \\ 
KO & 1 & -1.674 & 0.072 & 0.656 & 0.078 & 0.546 \\ 
 & 2 & 0.013 & 0.009 & 0.999 & 0.047 & 0.001 \\ 
MCD & 1 & -1.956 & 0.067 & 0.742 & 0.092 & 0.343 \\ 
 & 2 & -0.005 & 0.022 & 0.396 & 0.061 & 0 \\ 
MMM & 1 & 0.658 & 0.019 & 1 & 0.025 & 1 \\ 
 & 2 & -0.1 & 0.014 & 0.882 & 0.054 & 0 \\ 
MRK & 1 & 2.367 & 0.156 & 0.013 & 0.157 & 0.013 \\ 
 & 2 & 0.076 & 0.013 & 0.928 & 0.046 & 0.002 \\ 
MSFT & 1 & -2.175 & 0.176 & 0.003 & 0.165 & 0.007 \\ 
 & 2 & 0.26 & 0.016 & 0.753 & 0.057 & 0 \\ 
NKE & 1 & 1.652 & 0.05 & 0.956 & 0.1 & 0.258 \\ 
 & 2 & 0.079 & 0.022 & 0.416 & 0.059 & 0 \\ 
PG & 1 & 1.035 & 0.03 & 1 & 0.062 & 0.815 \\ 
 & 2 & 0.12 & 0.015 & 0.824 & 0.055 & 0 \\ 
TLT & 1 & -0.283 & 0.056 & 0.897 & 0.042 & 0.992 \\ 
 & 2 & -0.161 & 0.036 & 0.025 & 0.028 & 0.144 \\ 
TMF & 1 & -0.273 & 0.055 & 0.907 & 0.041 & 0.993 \\ 
 & 2 & -0.152 & 0.035 & 0.036 & 0.028 & 0.135 \\ 
TRV & 1 & -1.532 & 0.055 & 0.91 & 0.058 & 0.864 \\ 
 & 2 & -0.033 & 0.011 & 0.988 & 0.048 & 0.001 \\ 
UGL & 1 & 1.139 & 0.066 & 0.749 & 0.068 & 0.721 \\ 
 & 2 & -0.074 & 0.012 & 0.973 & 0.051 & 0 \\ 
UNH & 1 & 1.543 & 0.05 & 0.951 & 0.098 & 0.274 \\ 
 & 2 & 0.054 & 0.012 & 0.962 & 0.051 & 0 \\ 
V & 1 & 1.45 & 0.044 & 0.985 & 0.05 & 0.955 \\ 
 & 2 & 0.016 & 0.025 & 0.26 & 0.061 & 0 \\ 
VZ & 1 & -0.406 & 0.115 & 0.131 & 0.109 & 0.17 \\ 
 & 2 & -0.043 & 0.017 & 0.716 & 0.033 & 0.052 \\ 
WBA & 1 & 1.011 & 0.026 & 1 & 0.048 & 0.966 \\ 
 & 2 & 0.166 & 0.022 & 0.368 & 0.062 & 0 \\ 
WMT & 1 & 1.804 & 0.08 & 0.52 & 0.076 & 0.582 \\ 
 & 2 & -0.047 & 0.016 & 0.8 & 0.054 & 0 \\ 
\hline \\[-1.8ex] 
\end{tabular} 
\begin{tablenotes}
\item For NTS distribution, there are 5 $p$-values smaller than $0.05$. For $t$-distribution, there are $31$ $p$-values smaller than $0.05$. Most of the rejections correspond to Regime $2$, indicating that the marginal $t$-distribution is outperformed by marginal NTS distribution in the more volatile regime.  
\end{tablenotes}
\end{threeparttable}
\end{table} 

 For NTS distribution, there are 5 $p$-values smaller than $0.05$. For $t$-distribution, there are $31$ $p$-values smaller than $0.05$. Most of the rejections correspond to Regime $2$, indicating that the marginal $t$-distribution is outperformed by marginal NTS distribution in describing the return in the more volatile regime. We can also observe that $\beta$ varies significantly in different regimes. Generally, for all assets, $\beta$ in regime $1$ has larger absolute values.

\subsubsection{Transition Matrix and Denoised Correlation Matrices of Joint Innovation}

The transition matrix of the innovations is presented in Table \ref{transition}. The self-transition probability in two regimes are close, and both are high.

\begin{table}[!htbp] \centering 
\caption{Transition matrix of joint innovation} 
  \label{transition} 
\begin{threeparttable}

\begin{tabular}{@{\extracolsep{5pt}} c|cc} 

\hline \\[-1.8ex] 
 & regime $1$ & regime $2$ \\ 
\hline \\[-1.8ex] 
regime $1$ &0.8538&   0.1462\\ 
regime $2$ & 0.0301&  0.9699 \\
\hline
\end{tabular} 
\begin{tablenotes}
\item This table presents the transition matrix of the joint innovation of MRS-MNTS-GARCH model.
\end{tablenotes}
\end{threeparttable}
\end{table} 

We also provide the denoised correlation matrices of stdMNTS innovations in two~regimes in Table \ref{correlation}. Since there are two regimes identified in this period, we combine the two correlation matrices for easier comparison. The upper triangle of the matrix is the upper triangle of correlation matrix in regime $1$, while the lower triangle of the matrix is the lower triangle of correlation matrix in regime $2$. We find that the innovations have distinctively different correlation matrices in two regimes, validating the regime-switching assumption. The innovations are highly correlated in regime $2$, which corresponds to more volatile periods.

\subsection{Out-of-Sample Tests}
We conduct out-of-sample tests to demonstrate the outperformance of the optimal portfolios. We use 3 types of risk measures with different confidence levels in the optimization, namely, maximum drawdown, 0.7-CDaR, 0.3-CDaR, average drawdown, 0.5-CVaR, 0.7-CVaR, 0.9-CVaR and standard deviation. Note that 0-CVaR is equal to the expected return, which does not make sense as a risk measure. Minimizing $\alpha$-CVaR with small $\alpha$ means that we include the positive part of the return distribution as risk. This is a drawback of standard deviation, and thus we set $\alpha$ at 0.5, 0.7, and 0.9 to demonstrate the superiority. 

The rolling window technique is employed with $10$-day forward-moving time window. In each time window, we simulate 10,000 sample paths of length $10$ with the fitted model. The simulation is used as input to portfolio optimization. The portfolio is held for $10$~trading days before rebalancing. The portfolio optimization is performed with software Portfolio Safeguard (PSG) with precoded numerical optimization procedures.

\begin{sidewaystable*}[!htbp] \centering 
  \caption{Denoised correlation matrices of joint innovation in two regimes} 
  \label{correlation}
\setlength\tabcolsep{-2pt}

\begin{threeparttable}
 \tiny
\begin{tabular}{@{\extracolsep{5pt}} cccccccccccccccccccccccccccccccccc} 
\\[-1.8ex]\hline 
\\[-1.8ex] 
 & AAPL & AMGN & AXP & BA & CAT & CRM & CSCO & CVX & DIS & GS & HD & HON & IBM & INTC & JNJ & JPM & KO & MCD & MMM & MRK & MSFT & NKE & PG & SDOW & TMF & TRV & UGL & UNH & V & VZ & WBA & WMT \\ 
AAPL & $1$ & $0.15$ & $0.14$ & $0.09$ & $0.17$ & $0.23$ & $0.21$ & $0.09$ & $0.14$ & $0.12$ & $0.15$ & $0.25$ & $0.16$ & $0.13$ & $0.28$ & $0.08$ & $0.10$ & $0.04$ & $0.20$ & $0.11$ & $0.19$ & $0.23$ & $0.07$ & $$-$0.31$ & $$-$0.05$ & $0.08$ & $$-$0.01$ & $0.04$ & $0.34$ & $0.22$ & $0.21$ & $0.21$ \\ 
AMGN & $0.28$ & $1$ & $0.15$ & $0.11$ & $0.13$ & $0.40$ & $0.18$ & $0.12$ & $0.27$ & $0.16$ & $0.27$ & $0.29$ & $0.20$ & $0.18$ & $0.30$ & $0.16$ & $0.02$ & $0.06$ & $0.24$ & $0.39$ & $0.13$ & $0.29$ & $0.13$ & $$-$0.33$ & $$-$0.22$ & $$-$0.05$ & $$-$0.02$ & $0.04$ & $0.31$ & $0.13$ & $0.25$ & $0.10$ \\ 
AXP & $0.34$ & $0.39$ & $1$ & $0.37$ & $0.44$ & $0.15$ & $0.29$ & $0.33$ & $0.38$ & $0.50$ & $0.21$ & $0.38$ & $0.13$ & $0.13$ & $0.22$ & $0.48$ & $0.04$ & $0.15$ & $0.29$ & $0.28$ & $0.01$ & $0.27$ & $0.09$ & $$-$0.55$ & $$-$0.23$ & $0.18$ & $$-$0.16$ & $0.30$ & $0.16$ & $0.15$ & $0.21$ & $0.16$ \\ 
BA & $0.36$ & $0.40$ & $0.50$ & $1$ & $0.33$ & $0.09$ & $0.24$ & $0.13$ & $0.25$ & $0.28$ & $0.09$ & $0.35$ & $0.13$ & $0.05$ & $0.22$ & $0.28$ & $0.08$ & $0.13$ & $0.27$ & $0.18$ & $$-$0.01$ & $0.20$ & $0.08$ & $$-$0.44$ & $$-$0.14$ & $0.13$ & $$-$0.09$ & $0.26$ & $0.06$ & $0.17$ & $0.19$ & $0.12$ \\ 
CAT & $0.38$ & $0.32$ & $0.48$ & $0.49$ & $1$ & $0.18$ & $0.30$ & $0.30$ & $0.26$ & $0.36$ & $0.15$ & $0.43$ & $0.16$ & $0.28$ & $0.10$ & $0.33$ & $0.04$ & $0.06$ & $0.30$ & $0.15$ & $0.16$ & $0.11$ & $0.10$ & $$-$0.49$ & $$-$0.16$ & $0.14$ & $$-$0.04$ & $0.17$ & $0.11$ & $0.23$ & $0.10$ & $0.21$ \\ 
CRM & $0.31$ & $0.30$ & $0.37$ & $0.34$ & $0.36$ & $1$ & $0.23$ & $0.14$ & $0.36$ & $0.18$ & $0.32$ & $0.29$ & $0.29$ & $0.24$ & $0.23$ & $0.20$ & $0.09$ & $0.17$ & $0.30$ & $0.22$ & $0.28$ & $0.32$ & $0.07$ & $$-$0.33$ & $$-$0.18$ & $0.003$ & $$-$0.10$ & $$-$0.14$ & $0.39$ & $0.15$ & $0.23$ & $$-$0.03$ \\ 
CSCO & $0.35$ & $0.28$ & $0.37$ & $0.39$ & $0.40$ & $0.38$ & $1$ & $0.27$ & $0.36$ & $0.27$ & $$-$0.02$ & $0.30$ & $0.25$ & $0.28$ & $0.19$ & $0.29$ & $0.16$ & $0.13$ & $0.30$ & $0.21$ & $0.24$ & $0.09$ & $0.16$ & $$-$0.42$ & $$-$0.15$ & $0.12$ & $$-$0.09$ & $0.10$ & $0.22$ & $0.19$ & $0.19$ & $0.24$ \\ 
CVX & $0.32$ & $0.34$ & $0.46$ & $0.46$ & $0.58$ & $0.31$ & $0.38$ & $1$ & $0.22$ & $0.30$ & $0.10$ & $0.27$ & $0.17$ & $0.22$ & $0.22$ & $0.32$ & $0.14$ & $0.08$ & $0.34$ & $0.13$ & $0.11$ & $0.16$ & $0.17$ & $$-$0.45$ & $$-$0.30$ & $0.14$ & $$-$0.04$ & $0.13$ & $0.11$ & $0.20$ & $0.19$ & $0.17$ \\ 
DIS & $0.33$ & $0.39$ & $0.49$ & $0.49$ & $0.44$ & $0.37$ & $0.41$ & $0.44$ & $1$ & $0.32$ & $0.29$ & $0.40$ & $0.26$ & $0.09$ & $0.25$ & $0.32$ & $0.11$ & $0.19$ & $0.34$ & $0.08$ & $0.16$ & $0.34$ & $0.19$ & $$-$0.48$ & $$-$0.29$ & $0.13$ & $$-$0.17$ & $0.11$ & $0.34$ & $0.18$ & $0.21$ & $0.10$ \\ 
GS & $0.34$ & $0.39$ & $0.55$ & $0.48$ & $0.50$ & $0.35$ & $0.41$ & $0.47$ & $0.49$ & $1$ & $0.16$ & $0.31$ & $0.09$ & $0.21$ & $0.10$ & $0.64$ & $$-$0.08$ & $0.01$ & $0.26$ & $0.12$ & $0.09$ & $0.25$ & $$-$0.01$ & $$-$0.49$ & $$-$0.34$ & $0.10$ & $$-$0.30$ & $0.30$ & $0.20$ & $0.004$ & $0.16$ & $$-$0.04$ \\ 
HD & $0.34$ & $0.39$ & $0.48$ & $0.44$ & $0.41$ & $0.34$ & $0.36$ & $0.41$ & $0.48$ & $0.43$ & $1$ & $0.30$ & $0.27$ & $0.12$ & $0.12$ & $0.24$ & $$-$0.04$ & $0.14$ & $0.29$ & $0.13$ & $0.18$ & $0.29$ & $0.04$ & $$-$0.38$ & $$-$0.23$ & $0.13$ & $$-$0.17$ & $0.05$ & $0.29$ & $0.21$ & $0.17$ & $0.17$ \\ 
HON & $0.41$ & $0.43$ & $0.56$ & $0.63$ & $0.64$ & $0.42$ & $0.46$ & $0.57$ & $0.57$ & $0.55$ & $0.54$ & $1$ & $0.27$ & $0.22$ & $0.28$ & $0.37$ & $0.21$ & $0.24$ & $0.47$ & $0.17$ & $0.27$ & $0.28$ & $0.31$ & $$-$0.60$ & $$-$0.31$ & $0.31$ & $$-$0.18$ & $0.16$ & $0.33$ & $0.21$ & $0.36$ & $0.18$ \\ 
IBM & $0.36$ & $0.34$ & $0.43$ & $0.46$ & $0.46$ & $0.35$ & $0.44$ & $0.47$ & $0.43$ & $0.44$ & $0.41$ & $0.52$ & $1$ & $0.28$ & $0.03$ & $0.11$ & $0.08$ & $0.01$ & $0.30$ & $0.12$ & $0.30$ & $0.14$ & $0.14$ & $$-$0.39$ & $$-$0.13$ & $0.04$ & $$-$0.10$ & $$-$0.07$ & $0.31$ & $0.22$ & $0.14$ & $0.13$ \\ 
INTC & $0.37$ & $0.38$ & $0.44$ & $0.44$ & $0.46$ & $0.37$ & $0.45$ & $0.45$ & $0.43$ & $0.43$ & $0.41$ & $0.50$ & $0.48$ & $1$ & $0.06$ & $0.24$ & $0.10$ & $$-$0.04$ & $0.27$ & $0.17$ & $0.45$ & $0.08$ & $0.16$ & $$-$0.31$ & $$-$0.01$ & $$-$0.04$ & $0.01$ & $$-$0.02$ & $0.22$ & $0.21$ & $0.12$ & $0.09$ \\ 
JNJ & $0.28$ & $0.52$ & $0.45$ & $0.48$ & $0.41$ & $0.27$ & $0.37$ & $0.49$ & $0.45$ & $0.44$ & $0.44$ & $0.53$ & $0.45$ & $0.41$ & $1$ & $0.09$ & $0.18$ & $0.17$ & $0.33$ & $0.36$ & $0.04$ & $0.21$ & $0.25$ & $$-$0.37$ & $$-$0.06$ & $0.02$ & $0.04$ & $0.10$ & $0.34$ & $0.16$ & $0.21$ & $0.17$ \\ 
JPM & $0.35$ & $0.40$ & $0.59$ & $0.51$ & $0.52$ & $0.37$ & $0.43$ & $0.52$ & $0.52$ & $0.77$ & $0.46$ & $0.59$ & $0.47$ & $0.46$ & $0.49$ & $1$ & $$-$0.04$ & $0.09$ & $0.30$ & $0.15$ & $0.14$ & $0.24$ & $0.05$ & $$-$0.51$ & $$-$0.35$ & $0.24$ & $$-$0.31$ & $0.24$ & $0.20$ & $0.10$ & $0.18$ & $0.08$ \\ 
KO & $0.27$ & $0.35$ & $0.39$ & $0.42$ & $0.35$ & $0.28$ & $0.33$ & $0.46$ & $0.43$ & $0.33$ & $0.40$ & $0.47$ & $0.43$ & $0.40$ & $0.48$ & $0.40$ & $1$ & $0.23$ & $0.10$ & $0.13$ & $0.21$ & $0.06$ & $0.35$ & $$-$0.23$ & $$-$0.02$ & $0.40$ & $0.06$ & $0.05$ & $0.17$ & $0.26$ & $0.31$ & $0.14$ \\ 
MCD & $0.30$ & $0.35$ & $0.35$ & $0.39$ & $0.34$ & $0.33$ & $0.33$ & $0.38$ & $0.38$ & $0.33$ & $0.42$ & $0.42$ & $0.40$ & $0.35$ & $0.44$ & $0.35$ & $0.47$ & $1$ & $0.26$ & $0.09$ & $0.02$ & $0.15$ & $0.09$ & $$-$0.25$ & $$-$0.16$ & $0.35$ & $$-$0.14$ & $0.14$ & $$-$0.01$ & $0.02$ & $0.24$ & $0.07$ \\ 
MMM & $0.39$ & $0.42$ & $0.52$ & $0.56$ & $0.57$ & $0.37$ & $0.45$ & $0.55$ & $0.53$ & $0.51$ & $0.50$ & $0.67$ & $0.51$ & $0.50$ & $0.56$ & $0.55$ & $0.48$ & $0.45$ & $1$ & $0.17$ & $0.13$ & $0.31$ & $0.21$ & $$-$0.56$ & $$-$0.30$ & $0.29$ & $$-$0.20$ & $0.11$ & $0.27$ & $0.20$ & $0.23$ & $0.21$ \\ 
MRK & $0.24$ & $0.47$ & $0.40$ & $0.37$ & $0.35$ & $0.25$ & $0.32$ & $0.41$ & $0.38$ & $0.39$ & $0.35$ & $0.43$ & $0.37$ & $0.37$ & $0.52$ & $0.43$ & $0.37$ & $0.36$ & $0.44$ & $1$ & $0.05$ & $0.06$ & $0.18$ & $$-$0.36$ & $$-$0.06$ & $$-$0.03$ & $0.05$ & $0.27$ & $0.24$ & $0.16$ & $0.18$ & $0.22$ \\ 
MSFT & $0.37$ & $0.35$ & $0.43$ & $0.40$ & $0.43$ & $0.37$ & $0.46$ & $0.42$ & $0.42$ & $0.43$ & $0.40$ & $0.49$ & $0.48$ & $0.53$ & $0.40$ & $0.45$ & $0.39$ & $0.39$ & $0.47$ & $0.34$ & $1$ & $0.14$ & $0.17$ & $$-$0.28$ & $0.06$ & $0.03$ & $0.0002$ & $$-$0.04$ & $0.31$ & $0.29$ & $0.21$ & $0.12$ \\ 
NKE & $0.30$ & $0.32$ & $0.41$ & $0.41$ & $0.36$ & $0.38$ & $0.33$ & $0.33$ & $0.41$ & $0.36$ & $0.45$ & $0.48$ & $0.35$ & $0.35$ & $0.36$ & $0.39$ & $0.37$ & $0.43$ & $0.43$ & $0.31$ & $0.37$ & $1$ & $0.12$ & $$-$0.36$ & $$-$0.18$ & $0.12$ & $$-$0.13$ & $$-$0.06$ & $0.33$ & $0.07$ & $0.21$ & $0.14$ \\ 
PG & $0.27$ & $0.39$ & $0.36$ & $0.38$ & $0.31$ & $0.24$ & $0.33$ & $0.42$ & $0.40$ & $0.33$ & $0.38$ & $0.42$ & $0.40$ & $0.36$ & $0.52$ & $0.38$ & $0.54$ & $0.40$ & $0.47$ & $0.40$ & $0.38$ & $0.32$ & $1$ & $$-$0.32$ & $0.01$ & $0.15$ & $0.01$ & $0.08$ & $0.31$ & $0.19$ & $0.33$ & $0.21$ \\ 
SDOW & $$-$0.48$ & $$-$0.54$ & $$-$0.67$ & $$-$0.70$ & $$-$0.70$ & $$-$0.47$ & $$-$0.56$ & $$-$0.71$ & $$-$0.68$ & $$-$0.68$ & $$-$0.64$ & $$-$0.79$ & $$-$0.69$ & $$-$0.62$ & $$-$0.68$ & $$-$0.73$ & $$-$0.61$ & $$-$0.57$ & $$-$0.78$ & $$-$0.57$ & $$-$0.61$ & $$-$0.54$ & $$-$0.58$ & $1$ & $0.35$ & $$-$0.33$ & $0.20$ & $$-$0.34$ & $$-$0.42$ & $$-$0.34$ & $$-$0.35$ & $$-$0.30$ \\ 
TMF & $$-$0.27$ & $$-$0.24$ & $$-$0.38$ & $$-$0.33$ & $$-$0.37$ & $$-$0.27$ & $$-$0.28$ & $$-$0.34$ & $$-$0.36$ & $$-$0.42$ & $$-$0.31$ & $$-$0.38$ & $$-$0.32$ & $$-$0.30$ & $$-$0.32$ & $$-$0.45$ & $$-$0.25$ & $$-$0.24$ & $$-$0.37$ & $$-$0.26$ & $$-$0.28$ & $$-$0.29$ & $$-$0.23$ & $0.45$ & $1$ & $$-$0.27$ & $0.41$ & $$-$0.10$ & $$-$0.18$ & $$-$0.01$ & $$-$0.10$ & $0.03$ \\ 
TRV & $0.30$ & $0.39$ & $0.46$ & $0.46$ & $0.45$ & $0.31$ & $0.38$ & $0.48$ & $0.49$ & $0.51$ & $0.45$ & $0.54$ & $0.45$ & $0.42$ & $0.50$ & $0.56$ & $0.48$ & $0.42$ & $0.56$ & $0.39$ & $0.41$ & $0.38$ & $0.47$ & $$-$0.68$ & $$-$0.30$ & $1$ & $$-$0.19$ & $0.16$ & $0.03$ & $0.11$ & $0.23$ & $0.15$ \\ 
UGL & $0.05$ & $$-$0.06$ & $$-$0.01$ & $$-$0.01$ & $0.10$ & $$-$0.03$ & $$-$0.01$ & $0.14$ & $$-$0.05$ & $$-$0.05$ & $$-$0.06$ & $0.01$ & $0.001$ & $$-$0.01$ & $$-$0.001$ & $$-$0.07$ & $0.01$ & $$-$0.01$ & $0.02$ & $$-$0.02$ & $$-$0.02$ & $$-$0.07$ & $0.03$ & $$-$0.02$ & $0.16$ & $$-$0.01$ & $1$ & $$-$0.05$ & $$-$0.22$ & $0.09$ & $$-$0.04$ & $0.04$ \\ 
UNH & $0.30$ & $0.40$ & $0.37$ & $0.38$ & $0.33$ & $0.26$ & $0.30$ & $0.38$ & $0.38$ & $0.39$ & $0.38$ & $0.44$ & $0.38$ & $0.33$ & $0.42$ & $0.42$ & $0.34$ & $0.31$ & $0.41$ & $0.36$ & $0.35$ & $0.33$ & $0.30$ & $$-$0.54$ & $$-$0.25$ & $0.41$ & $$-$0.002$ & $1$ & $$-$0.03$ & $0.09$ & $0.13$ & $0.07$ \\ 
V & $0.32$ & $0.37$ & $0.51$ & $0.43$ & $0.38$ & $0.36$ & $0.34$ & $0.39$ & $0.43$ & $0.44$ & $0.41$ & $0.49$ & $0.40$ & $0.37$ & $0.42$ & $0.46$ & $0.35$ & $0.36$ & $0.46$ & $0.35$ & $0.38$ & $0.39$ & $0.34$ & $$-$0.60$ & $$-$0.31$ & $0.43$ & $$-$0.02$ & $0.34$ & $1$ & $0.16$ & $0.17$ & $0.13$ \\ 
VZ & $0.26$ & $0.36$ & $0.39$ & $0.35$ & $0.36$ & $0.23$ & $0.30$ & $0.42$ & $0.40$ & $0.34$ & $0.37$ & $0.44$ & $0.38$ & $0.36$ & $0.47$ & $0.40$ & $0.46$ & $0.37$ & $0.46$ & $0.38$ & $0.36$ & $0.31$ & $0.43$ & $$-$0.56$ & $$-$0.21$ & $0.46$ & $0.01$ & $0.26$ & $0.33$ & $1$ & $0.18$ & $0.27$ \\ 
WBA & $0.25$ & $0.36$ & $0.33$ & $0.36$ & $0.31$ & $0.28$ & $0.26$ & $0.32$ & $0.34$ & $0.33$ & $0.35$ & $0.41$ & $0.33$ & $0.31$ & $0.41$ & $0.35$ & $0.33$ & $0.31$ & $0.38$ & $0.34$ & $0.33$ & $0.29$ & $0.36$ & $$-$0.48$ & $$-$0.22$ & $0.36$ & $$-$0.01$ & $0.32$ & $0.29$ & $0.30$ & $1$ & $0.19$ \\ 
WMT & $0.20$ & $0.30$ & $0.30$ & $0.32$ & $0.24$ & $0.18$ & $0.28$ & $0.28$ & $0.33$ & $0.28$ & $0.39$ & $0.37$ & $0.30$ & $0.27$ & $0.40$ & $0.32$ & $0.38$ & $0.34$ & $0.36$ & $0.32$ & $0.31$ & $0.30$ & $0.41$ & $$-$0.47$ & $$-$0.17$ & $0.38$ & $$-$0.02$ & $0.29$ & $0.27$ & $0.37$ & $0.30$ & $1$ \\  
\hline \\[-1.8ex] 
\end{tabular} 

\begin{tablenotes}
\item This table presents the denoised correlation matrices of joint innovations in two regimes. The upper triangle corresponds the regime $1$ while the lower triangle corresponds to regime $2$ which is more volatile. The correlations in regime $2$ are generally higher than those in regime $1$.
\end{tablenotes}
\end{threeparttable}
\end{sidewaystable*} 

\subsubsection{Optimal Portfolios}
We report the performance of the optimal portfolios in various forms. Due to a large number of risk measures, we sometimes only present representative ones in graphs for better visualization. The other optimal portfolios have similar results.

\paragraph{Performance Measures}
In Table \ref{performancetable}, we measure the performance of the optimal portfolios by various performance ratios. The row names are the risk measures used in portfolio optimization. The columns are the performance measures. For example, the 0.3-CDaR portfolio is optimized every $10$ days to maximize the ratio of expected cumulative return over 0.3-CDaR.  Note that 0-CDaR is the average drawdown while 1-CDaR is the maximum drawdown. The risk measure standard deviation is studied in the classical Markowitz model, of which the optimization is also called MV optimization.

We can observe that the optimal portfolios of CDaR and CVaR measures have very close performance regardless of performance measures. The confidence level has a marginal impact on the ratios. Considering all the ratios, 0.3-CDaR portfolio and 0.5-CVaR portfolio are the best among the ten portfolios. 0.3-CDaR portfolio outperforms the others in all return-CVaR ratios, while 0.5-CVaR portfolio outperforms the others in return-CDaR ratios, slightly but consistently. All of the optimal portfolios of CDaR and CVaR measures consistently outperform the MV optimal portfolio, which significantly outperforms the DJIA and the equally weighted portfolio.

   \begin{table}[!htbp] 
   \centering
    \caption{Performance ratios of  optimal portfolios with different risk measures} 
  \label{performancetable} 
  \begin{threeparttable} \centering
 \setlength\tabcolsep{-1pt}
 \scriptsize

\begin{tabular}{@{\extracolsep{5pt}} cccccccccc} 
\\[-1.8ex]
\hline \\[-1.8ex] 
 & $\frac{\text{Mean Return}}{0\text{-CDaR}}$ & $\frac{\text{Mean Return}}{0.3\text{-CDaR}}$ & $\frac{\text{Mean Return}}{0.7\text{-CDaR}}$ & $\frac{\text{Mean Return}}{1\text{-CDaR}}$& $\frac{\text{Mean Return}}{0.5\text{-CVaR}}$ & $\frac{\text{Mean Return}}{0.7\text{-CVaR}}$ & $\frac{\text{Mean Return}}{0.9\text{-CVaR}}$ & $\scriptsize{\text{Sharpe Ratio}}$ & $\scriptsize{\text{Rachev Ratio}}$\\[0.1cm]
\hline  \\  [-1.8ex] 
0-CDaR & $0.072$ & $0.051$ & $0.027$ & $0.008$ & $0.258$ & $0.156$ & $0.077$ & $0.117$ & $0.965$ \\ 
0.3-CDaR & $0.074$ & $0.052$ & $0.027$ & $0.008$ & $0.262$ & $0.159$ & $0.078$ & $0.119$ & $0.968$ \\ 
0.7-CDaR & $0.073$ & $0.052$ & $0.027$ & $0.008$ & $0.260$ & $0.158$ & $0.077$ & $0.118$ & $0.965$ \\ 
1-CDaR & $0.075$ & $0.053$ & $0.028$ & $0.007$ & $0.259$ & $0.157$ & $0.078$ & $0.117$ & $1.012$ \\ 
0.5-CVaR & $0.076$ & $0.054$ & $0.028$ & $0.007$ & $0.259$ & $0.158$ & $0.076$ & $0.120$ & $1.002$ \\ 
0.7-CVaR & $0.072$ & $0.051$ & $0.027$ & $0.007$ & $0.257$ & $0.157$ & $0.077$ & $0.121$ & $1.000$ \\ 
0.9-CVaR & $0.071$ & $0.050$ & $0.027$ & $0.007$ & $0.243$ & $0.147$ & $0.072$ & $0.116$ & $0.986$ \\ 
Standard Deviation & $0.064$ & $0.045$ & $0.024$ & $0.006$ & $0.238$ & $0.145$ & $0.071$ & $0.109$ & $0.966$ \\ 
DJIA & $0.011$ & $0.008$ & $0.004$ & $0.001$ & $0.064$ & $0.040$ & $0.019$ & $0.033$ & $0.855$ \\ 
Equal Weight & $0.032$ & $0.023$ & $0.011$ & $0.003$ & $0.136$ & $0.083$ & $0.037$ & $0.065$ & $0.900$ \\ 
\hline \\
\end{tabular}
\begin{tablenotes}
\item The column names are different  ratios used to measure the performance. The row names show the risk measures used in the portfolio optimization. The DJIA index and the equally weighed portfolio are included as benchmarks.
\end{tablenotes}
\end{threeparttable}
\end{table}

\paragraph{Log Cumulative Return}
 
\begin{figure*}[!htbp]
\centering

    \includegraphics[width=.80\textwidth]{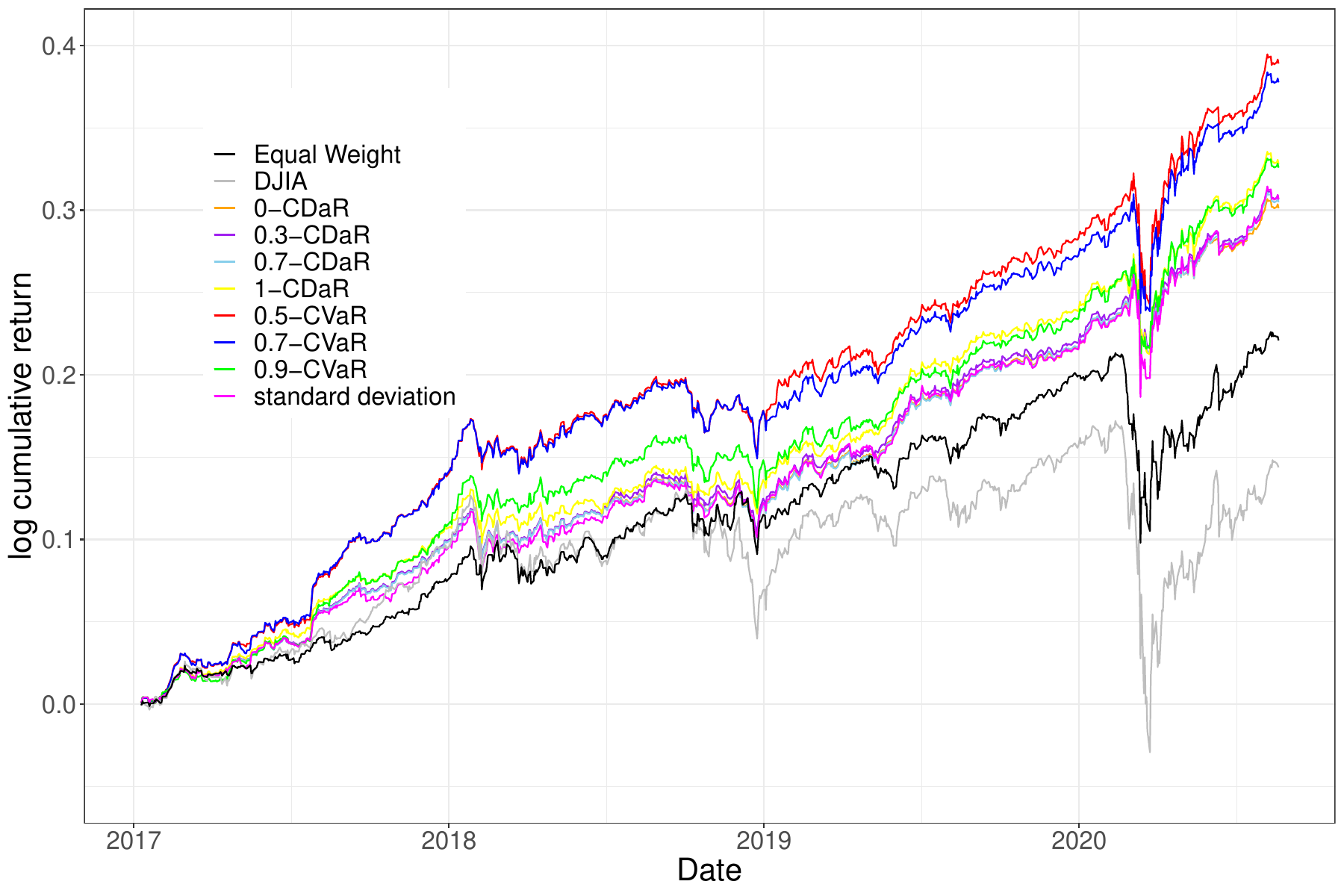}

  \caption{Log cumulative returns: this figure plots the time series of log cumulative return of optimal portfolios with different risk measures, the DJIA index and the equally weighted portfolio. The risk measures used in portfolio optimization are 0-CDaR, 0.3-CDaR, 0.7-CDaR, 1-CDaR, 0.5-CVaR, 0.7-CVaR, 0.9-CVaR and standard deviation.}\
  \label{performancegraph}
\end{figure*}

In Figure \ref{performancegraph}, we plot the time series of log cumulative returns of the optimal portfolios with different risk measures in out-of-sample tests to visualize the performance. The legend shows colors that represent different risk measures used in each portfolio optimization. We use log base 10 compounded cumulative returns as a vertical axis so that the scale of relative drawdown can be easily compared in graphs by counting the grids. The performance with different risk measures is reported separately. The labels in the legend show the confidence level of the risk measure. The names of the subfigures indicate the risk measures used in the optimization. DJIA index and the equally weighted portfolio are included in each graph for comparison.

As can be observed in the graphs, all the optimal portfolios follow a similar trend, though they differ in overall performance. They tend to alleviate both left tail and right tail events. CVaR optimal portfolios have higher cumulative returns than CDaR optimal portfolios. The confidence level has only a marginal impact. CVaR optimal portfolios with confidence levels 0.5 and 0.7 are almost indistinguishable. CDaR optimal portfolios with confidence levels 0.5, 0.7 and 0.9 are almost indistinguishable.

\paragraph{Relative Drawdown}

\begin{figure}[!htbp]
\centering
 
    \includegraphics[width=.80\textwidth]{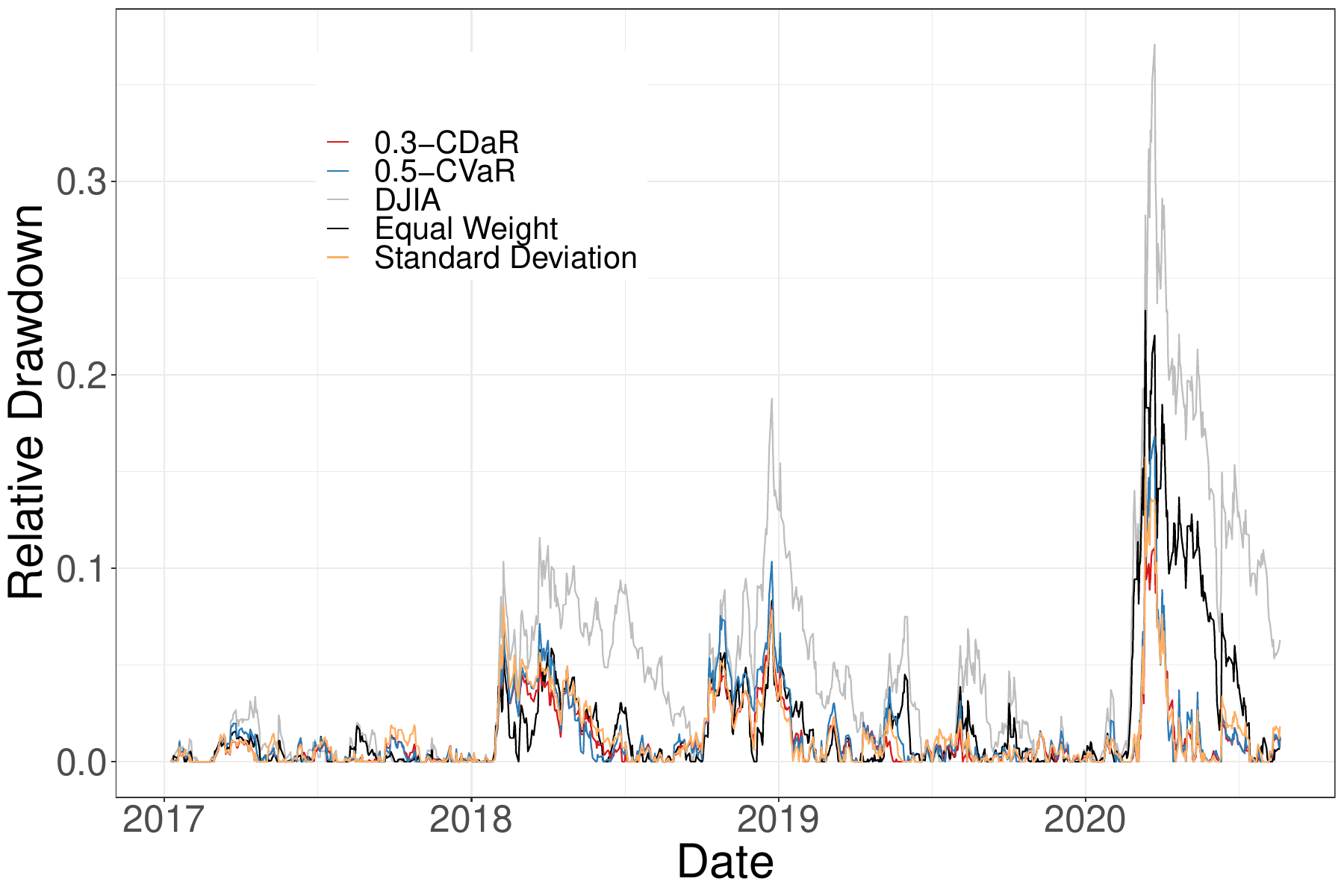}
 
  \caption{Relative drawdown:  this figure plots the relative drawdown paths of optimal portfolios with different risk measures, the DJIA index and the equally weighted portfolio. The risk measures include 0-CDaR, 0.5-CVaR and standard deviation.}
  \label{drawdown} 
\end{figure}

Since the CVaR and CDaR risk measures only concern the tail behavior of a portfolio,  we plot the relative drawdown  of the optimal portfolios in out-of-sample testing period in Figure \ref{drawdown} to demonstrate their ability to alleviate extreme left tail events.
Due to a large number of optimal portfolios with different risk measures, we only plot 0-CDaR, 0.5-CVaR and MV optimal portfolios here for better visualization. The others have similar results.
The 0-CDaR, 0.5-CVaR and MV optimal portfolios have significantly smaller relative drawdown most of the time in the out-of-sample test, especially in year 2020. 0-CDaR and 0.5-CVaR optimal portfolios are slightly better than the MV optimal portfolio. 

\paragraph{Portfolio Return Distribution}

\begin{figure}[!htbp]
\centering
	\subfloat[Kernel density estimation:  this figure plots the kernel density estimation of optimal portfolios with different risk measures, the DJIA index and the equally weighted portfolio. The risk measures include 0-CDaR, 0.5-CVaR and standard deviation.]{
    \includegraphics[width=.44\textwidth]{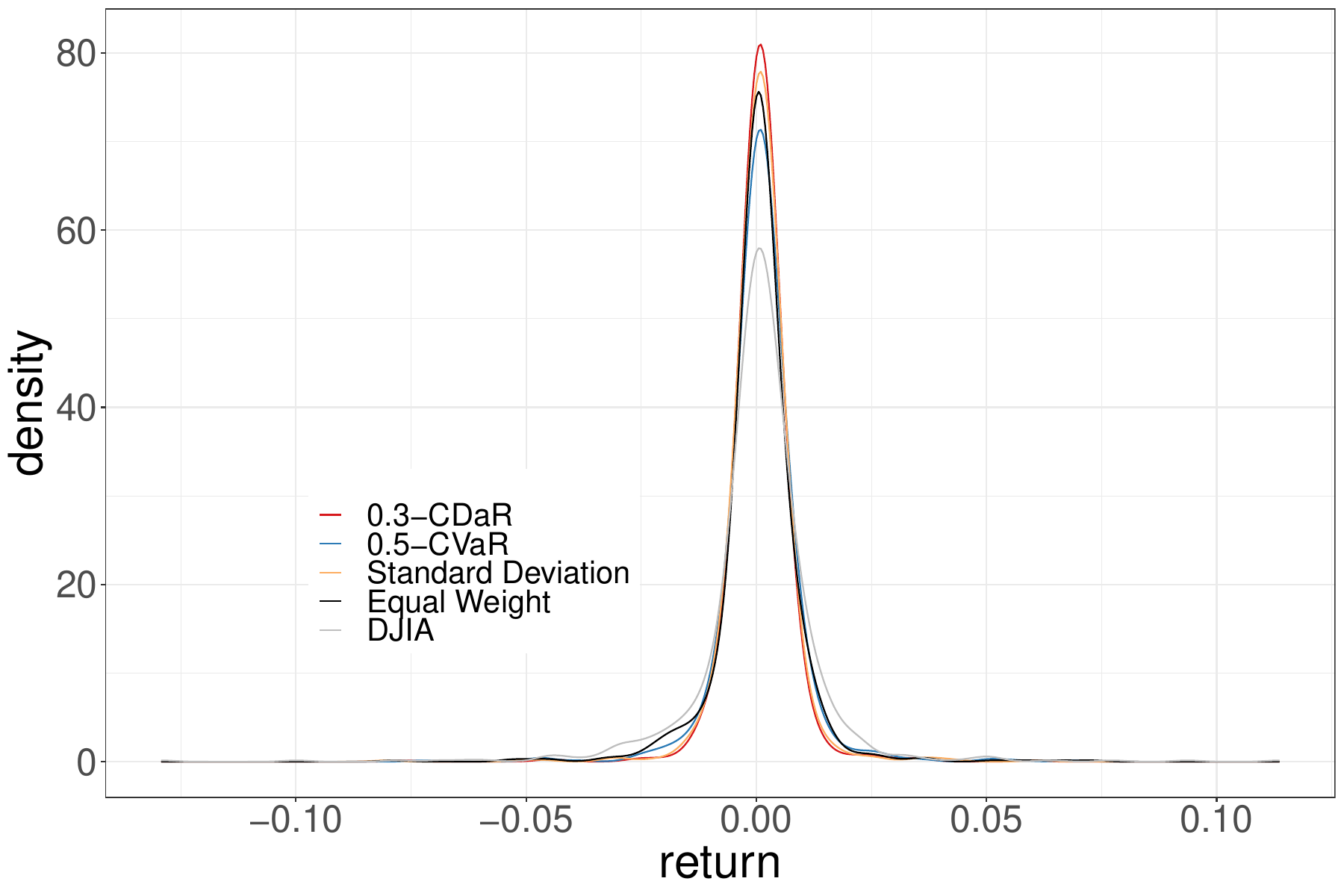}}
    \hfill  
 	\subfloat[log kernel density estimation:  this figure plots the log kernel density estimation of optimal portfolios with different risk measures, the DJIA index and the equally weighted portfolio. The risk measures include 0-CDaR, 0.5-CVaR and standard deviation.]{
   \includegraphics[width=.44\textwidth]{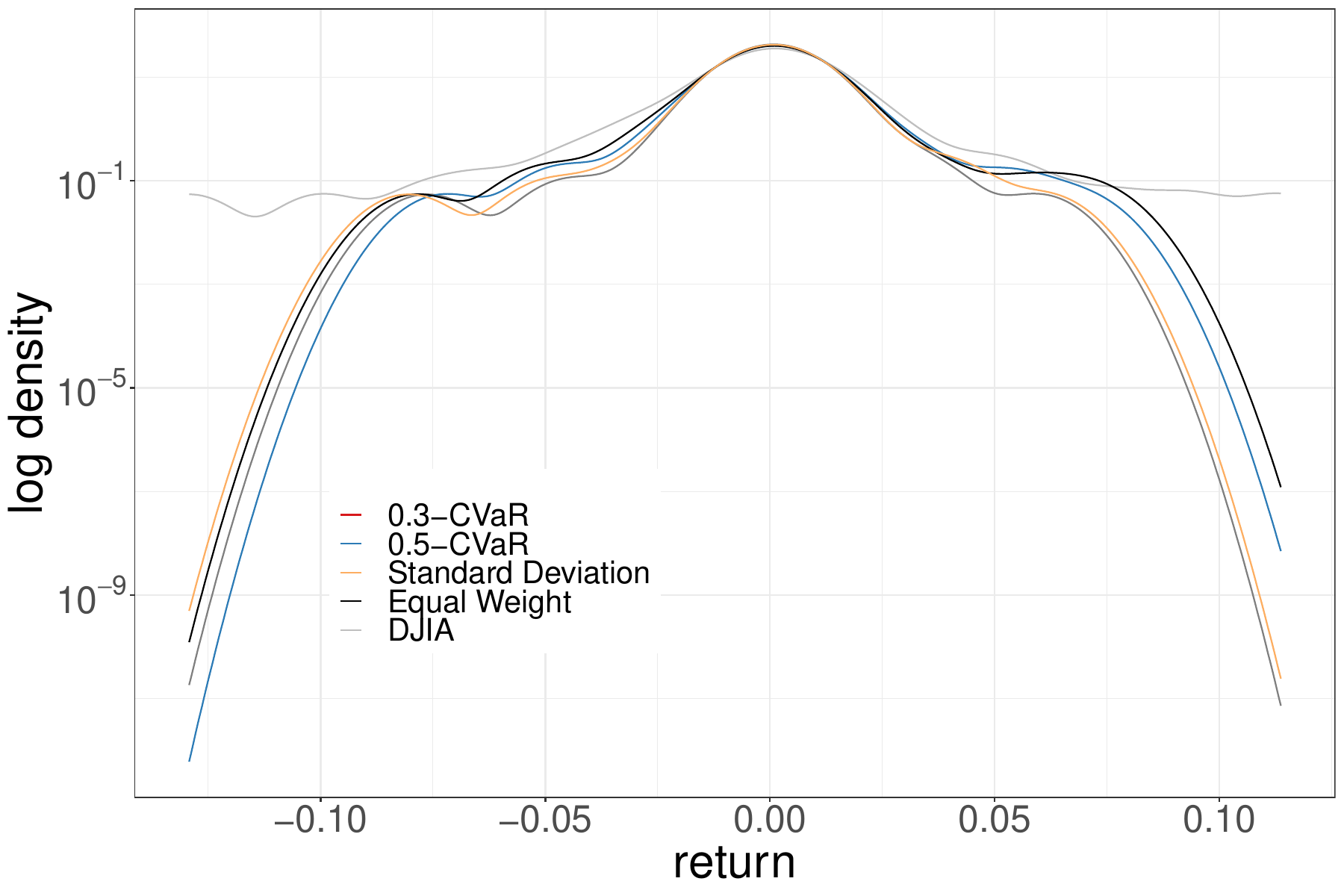}}
  \caption{Kernel density estimation and log kernel density estimation}\label{density}  
\end{figure}

In Figure \ref{density}, we plot the kernel density estimation on the return distribution. To better compare the tails of the return distribution, we also plot the log base 10 density. We can easily observe that the optimal portfolios are very close and alleviate both right and left tail events. Unlike standard deviation (MV) optimal portfolio, 0.3-CDaR and 0.5-CVaR portfolio have thinner tails on both sides than DJIA and the equally weighted portfolio. 0.3-CDaR and standard deviation portfolio have a higher peak than the equally weighted portfolio. Among the three optimal portfolios, the 0.5-CVaR portfolio has the thinnest left tail and a fatter right tail than the other two optimal portfolios. 

\paragraph{Allocation}

\begin{figure}[!htbp]
\centering
\includegraphics[width=.80\textwidth]{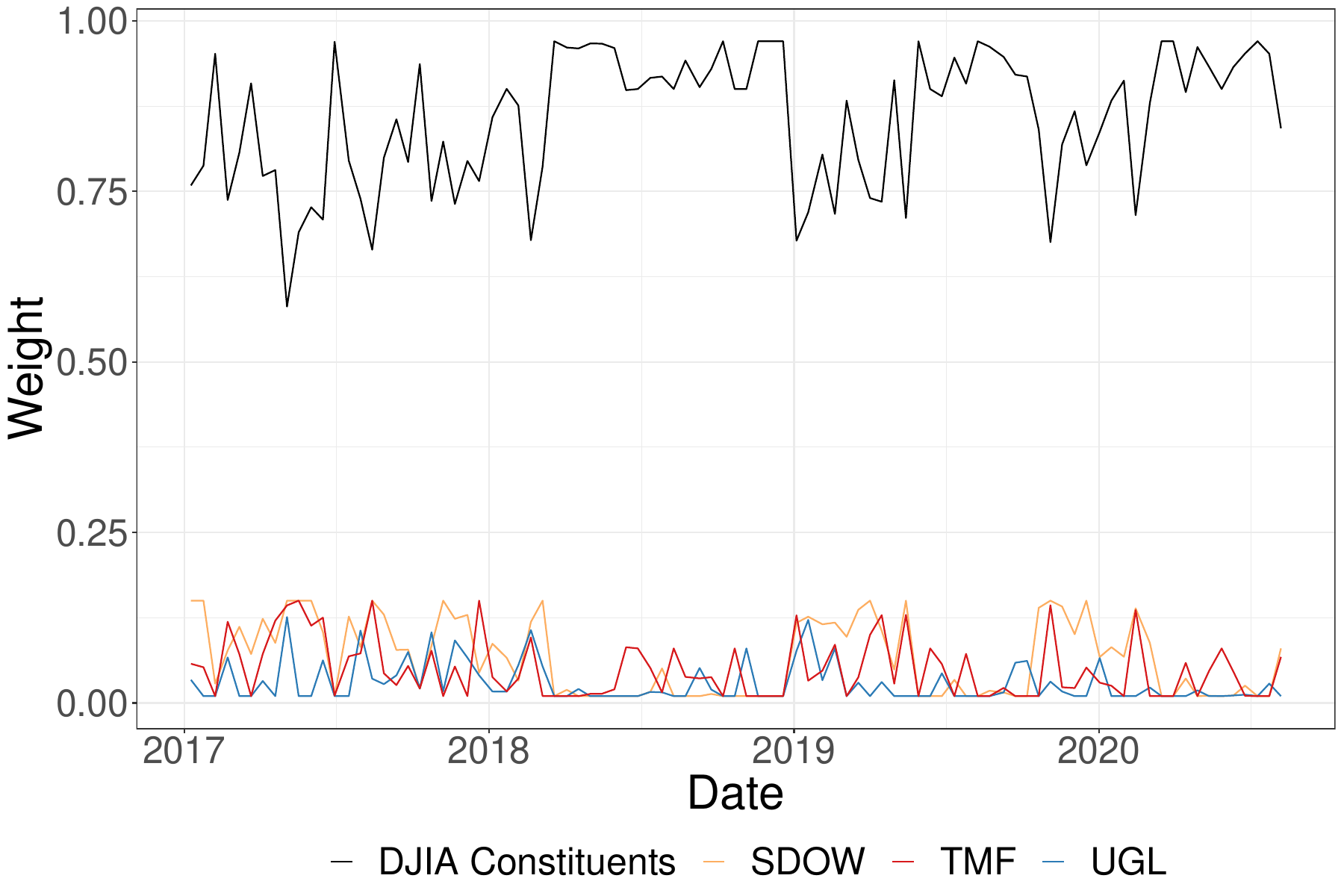}
      \caption{Allocation: this figure plots the weights of 0-CDaR optimal portfolio on DJIA constituents and 3 ETFs. }
      \label{weights} 
\end{figure}

In Figure \ref{weights}, we report the time series of optimal weights of the 0-CDaR optimal portfolio as a representative case. The DJIA constituents are aggregated for better visualization. It is reasonable that when the total weight on DJIA constituents is high, the weights on shortselling ETF is low. The weights on ETFs hit the lower and upper bounds 0.01 and 0.15 in many periods, indicating that the constraints on weights are active in the optimizations.

\subsubsection{Robustness to Suboptimality in Optimization}

So far, we have been studying the optimal portfolios that maximize performance ratios. In this section, we study the sensitivity of the performance by comparing suboptimal portfolios with the optimal ones. We show that the portfolio with tail risk measure is more robust to suboptimality in optimization than the mean-variance portfolio. The performance ratios of CDaR and CVaR optimal portfolios are shown to be not sensitive to  changing the value of return constraint, while the performance ratios of mean-variance optimal portfolio shows significant change. The graphs of log cumulative return show that the suboptimal portfolios follow a similar trend, sometimes overlapping with each other.

We use 9 portfolios with different constraints on return to approximate the efficient frontier for each risk measure. For example, in each 0.3-CDaR optimization, we perform 10 optimizations with varying constraints on return. The portfolio with the highest return-risk ratio is the approximated optimal portfolio. We denote the portfolio that has $n$ lower level of constraint on return as level L$n$ suboptimal portfolio, the one that has $n$ higher level of constraint on return as level H$n$ suboptimal portfolio. When a certain optimization is unfeasible, we set the allocation the same as that of a lower level. The constraints on return are  (0.002, 0.010, 0.020, 0.030, 0.035, 0.040, 0.045, 0.050, 0.060, 0.065).
For example, when the portfolio with constraint 0.050 on return has the highest performance ratio among the 10~portfolios, it is labeled as the optimal portfolio. Accordingly, the portfolio with one-level higher constraint 0.060 is labeled H1. In this case, H2 to H4 have the same constraint 0.065, since there is no optimization performed with higher constraint. Note that at each rebalance, the optimal portfolio changes, so do portfolios of other levels, i.e., constraint 0.060 may no longer be optimal. The performance of each label is measured with portfolios of corresponding level determined at each rebalance.

\begin{figure*}[!htbp]
\centering
  \subfloat[0-CDaR Suboptimal portfolios: this figure plots the paths of 0-CDaR suboptimal portfolios from level H1 to L4, and the equally weighted portfolio.]{%
    \includegraphics[width=.44\textwidth]{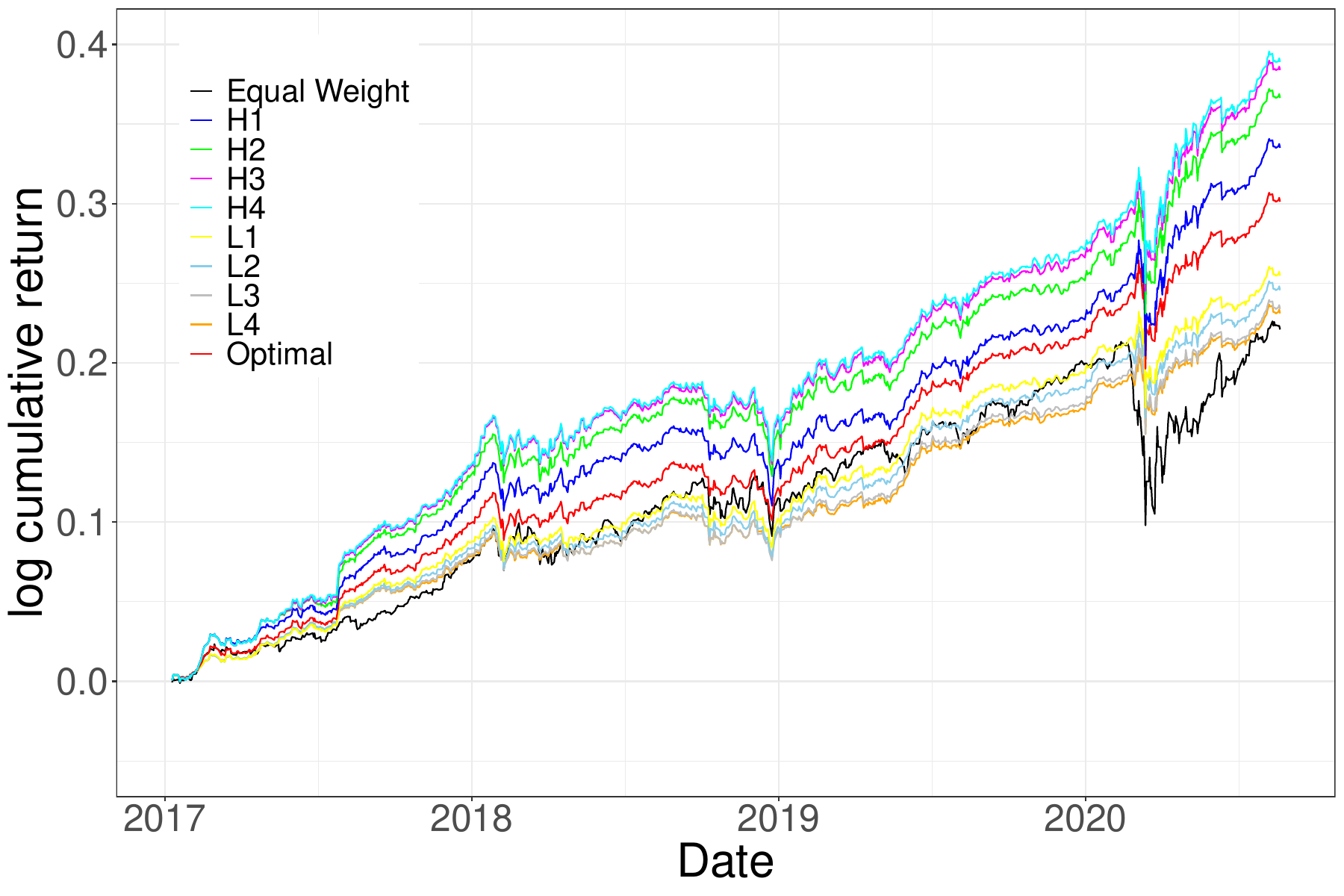}}\\
    \vspace{0.4cm}
    \subfloat[ 0.5-CVaR Suboptimal portfolios:  this figure plots the paths of 0.5-CVaR suboptimal portfolios from level H1 to L4, and the equally weighted portfolio.]{%
    \includegraphics[width=.44\textwidth]{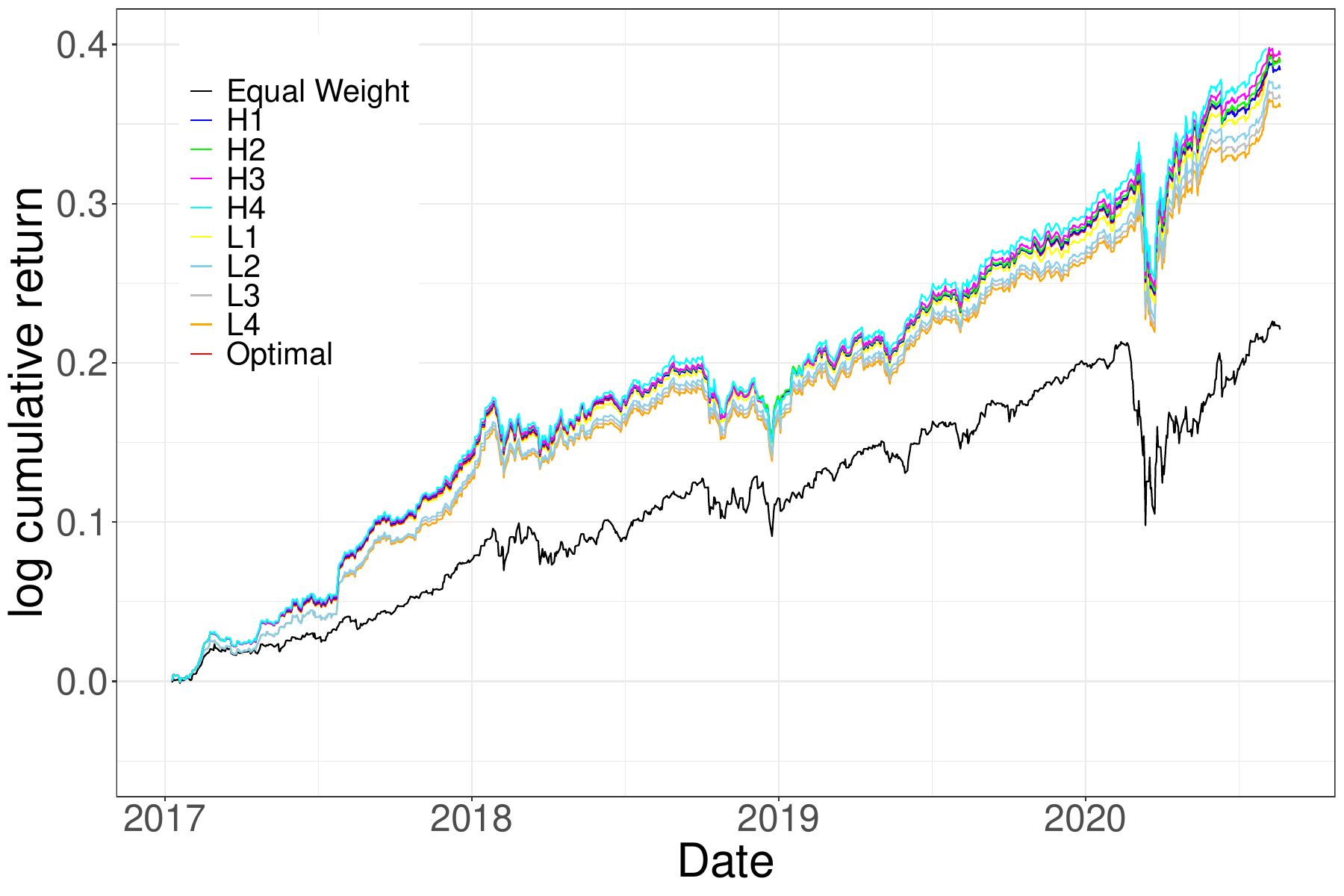}}\hfill
     \subfloat[MV suboptimal portfolios:  this figure plots the paths of mean variance suboptimal portfolios from level H1 to L4, and the equally weighted portfolio.]{%
    \includegraphics[width=.44\textwidth]{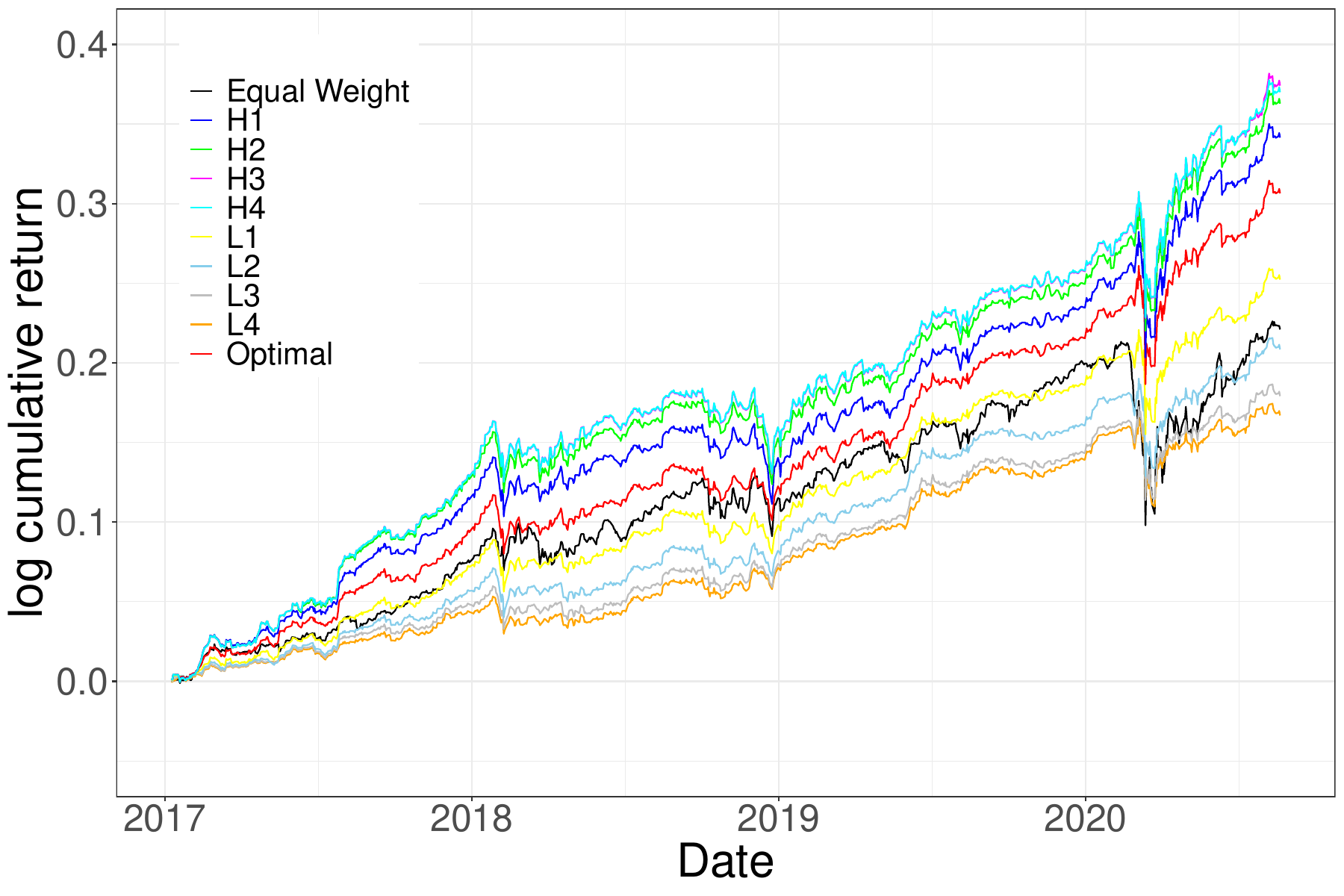}}
 
  \caption{Time series of log cumulative return of suboptimal portfolios with different risk measures}\
  \label{suboptperformancegraph}
\end{figure*}

We report suboptimal portfolios with risk measures of 0-CDaR, 0.5-CVaR and standard deviation in Figure \ref{suboptperformancegraph} and Table \ref{suboptperformancetable} as representative cases. We find that the optimal portfolio is consistently too conservative for all risk measures and input. That is, with the same risk measure, suboptimal portfolios with a higher constraint on return achieve higher performance ratios than the optimal portfolio. From $H4$ to $L4$, the ratios do not show a pattern of first increasing the decreasing as expected. In other words, the out-of-sample efficient frontiers are usually non-convex with irregular shapes. Overall, the performance ratios of CDaR and CVaR optimal portfolios are shown to be not sensitive to  changing the value of return constraint, while the performance ratios of mean-variance optimal portfolio shows significant change.

\begin{table}[!htbp] \centering 
 \caption{ Performance ratios of suboptimal portfolios with different risk measures} 
  \label{suboptperformancetable}
\begin{threeparttable}
\setlength\tabcolsep{-1pt}
\scriptsize
\begin{tabular}{@{\extracolsep{5pt}} cccccccccc} 
\\[-1.8ex]\hline 
 \\[-1.8ex] 
 & $\frac{\text{Mean Return}}{0\text{-CDaR}}$ & $\frac{\text{Mean Return}}{0.3\text{-CDaR}}$ & $\frac{\text{Mean Return}}{0.7\text{-CDaR}}$ & $\frac{\text{Mean Return}}{1\text{-CDaR}}$& $\frac{\text{Mean Return}}{0.5\text{-CVaR}}$ & $\frac{\text{Mean Return}}{0.7\text{-CVaR}}$ & $\frac{\text{Mean Return}}{0.9\text{-CVaR}}$ & $\small{\text{Sharpe Ratio}}$ & $\small{\text{Rachev Ratio}}$\\[0.1cm]
\hline \\[-1.8ex] 
0-CDaR\\
L4 & $0.061$ & $0.043$ & $0.023$ & $0.008$ & $0.239$ & $0.146$ & $0.074$ & $0.113$ & $0.965$ \\ 
L3 & $0.058$ & $0.041$ & $0.021$ & $0.007$ & $0.232$ & $0.142$ & $0.071$ & $0.108$ & $0.959$ \\ 
L2 & $0.061$ & $0.043$ & $0.023$ & $0.007$ & $0.236$ & $0.144$ & $0.072$ & $0.110$ & $0.960$ \\ 
L1 & $0.062$ & $0.044$ & $0.023$ & $0.007$ & $0.236$ & $0.144$ & $0.072$ & $0.110$ & $0.959$ \\ 
optimal & $0.072$ & $0.051$ & $0.027$ & $0.008$ & $0.258$ & $0.156$ & $0.077$ & $0.117$ & $0.965$ \\ 
H1 & $0.074$ & $0.052$ & $0.028$ & $0.008$ & $0.251$ & $0.151$ & $0.073$ & $0.115$ & $0.968$ \\ 
H2 & $0.082$ & $0.058$ & $0.030$ & $0.008$ & $0.259$ & $0.157$ & $0.076$ & $0.120$ & $0.993$ \\ 
H3 & $0.087$ & $0.061$ & $0.032$ & $0.009$ & $0.267$ & $0.161$ & $0.078$ & $0.124$ & $0.995$ \\ 
H4 & $0.087$ & $0.062$ & $0.032$ & $0.009$ & $0.268$ & $0.162$ & $0.079$ & $0.125$ & $0.989$ \\ \hline
0.5-CVaR\\
L4 & $0.073$ & $0.052$ & $0.027$ & $0.006$ & $0.252$ & $0.154$ & $0.074$ & $0.114$ & $1.011$ \\ 
L3 & $0.074$ & $0.052$ & $0.027$ & $0.007$ & $0.253$ & $0.155$ & $0.075$ & $0.115$ & $1.011$ \\ 
L2 & $0.074$ & $0.052$ & $0.027$ & $0.007$ & $0.253$ & $0.155$ & $0.075$ & $0.116$ & $1.008$ \\ 
L1 & $0.075$ & $0.053$ & $0.027$ & $0.007$ & $0.256$ & $0.156$ & $0.076$ & $0.118$ & $1.006$ \\ 
optimal & $0.076$ & $0.054$ & $0.028$ & $0.007$ & $0.259$ & $0.158$ & $0.076$ & $0.120$ & $1.002$ \\ 
H1 & $0.075$ & $0.053$ & $0.027$ & $0.007$ & $0.255$ & $0.155$ & $0.075$ & $0.118$ & $0.990$ \\ 
H2 & $0.075$ & $0.053$ & $0.027$ & $0.007$ & $0.256$ & $0.156$ & $0.075$ & $0.119$ & $0.987$ \\ 
H3 & $0.073$ & $0.052$ & $0.027$ & $0.007$ & $0.256$ & $0.155$ & $0.075$ & $0.119$ & $0.982$ \\ 
H4 & $0.071$ & $0.050$ & $0.026$ & $0.007$ & $0.254$ & $0.154$ & $0.074$ & $0.117$ & $0.977$ \\ \hline
Standard Deviation \\
L4 & $0.041$ & $0.029$ & $0.015$ & $0.004$ & $0.176$ & $0.110$ & $0.055$ & $0.084$ & $0.990$ \\ 
L3 & $0.042$ & $0.030$ & $0.016$ & $0.004$ & $0.177$ & $0.111$ & $0.056$ & $0.083$ & $1.005$ \\ 
L2 & $0.049$ & $0.035$ & $0.019$ & $0.005$ & $0.196$ & $0.122$ & $0.062$ & $0.092$ & $1.013$ \\ 
L1 & $0.059$ & $0.042$ & $0.022$ & $0.006$ & $0.225$ & $0.138$ & $0.069$ & $0.103$ & $1.005$ \\ 
optimal & $0.064$ & $0.045$ & $0.024$ & $0.006$ & $0.238$ & $0.145$ & $0.071$ & $0.109$ & $0.966$ \\ 
H1 & $0.069$ & $0.049$ & $0.026$ & $0.007$ & $0.232$ & $0.141$ & $0.069$ & $0.108$ & $0.972$ \\ 
H2 & $0.071$ & $0.050$ & $0.027$ & $0.007$ & $0.232$ & $0.141$ & $0.069$ & $0.110$ & $0.983$ \\ 
H3 & $0.074$ & $0.052$ & $0.028$ & $0.007$ & $0.237$ & $0.144$ & $0.070$ & $0.113$ & $0.986$ \\ 
H4 & $0.073$ & $0.052$ & $0.027$ & $0.007$ & $0.234$ & $0.142$ & $0.069$ & $0.112$ & $0.980$ \\ \hline
DJIA & $0.011$ & $0.008$ & $0.004$ & $0.001$ & $0.064$ & $0.040$ & $0.019$ & $0.033$ & $0.855$ \\ 
Equal Weight & $0.032$ & $0.023$ & $0.011$ & $0.003$ & $0.136$ & $0.083$ & $0.037$ & $0.065$ & $0.900$ \\ 
\hline \\[-1.8ex] 
\end{tabular} 

\begin{tablenotes}
\item This table presents the performance ratios of suboptimal portfolios with different risk measures. The risk measures are shown in the row names: 0-CDaR, 0.5-CVaR and standard deviation. The constraint on return is adjusted from high to low (H4 to L4) to show the impact of suboptimality. Portfolio optimization with tail risk measures demonstrates smaller variation in out-of-sample performance ratios compared with mean-variance portfolio.
\end{tablenotes}
\end{threeparttable}
\end{table}

The portfolios have increasing returns as well as risk from L1 to H4. The realized paths follow a similar trend and rarely cross. The optimal portfolios, as expected, are in the medium part of all paths. The  0.5-CVaR suboptimal portfolios have almost identical performance. Some paths are not visible due to overlap. For example, the constraints on the return of H4 are sometimes not feasible, leading to the same performance with H3~portfolio.

\section{Conclusions}
We propose the MRS-MNTS-GARCH model that is sufficiently flexible to accommodate fat tails, skewness and regime switch in asset returns.  The model is used to simulate sample paths of portfolio returns, which serves as input to portfolio optimization with tail risk measures. We conduct various in-sample and out-of-sample tests to demonstrate the effectiveness of our approach. In-sample tests show that the NTS distribution fits the innovations with high accuracy compared with $t$-distribution. Out-of-sample tests show that our approach significantly improves the performance of optimal portfolios measured by performance ratios and it successfully mitigates left tail risk. 
We also find that
out-of-sample performance ratios of the portfolios with tail risk measures are more robust to suboptimality 
on the efficient frontier compared with mean-variance portfolio.

\bibliographystyle{unsrtnat}
\bibliography{Bibliography}

\end{document}